# Large-Block Modular Addition Checksum Algorithms

Philip Koopman
Carnegie Mellon University
Pittsburgh, PA, USA
koopman@cmu.edu

***Abstract*—Checksum algorithms are widely employed due to their use of a simple algorithm with fast computational speed to provide a basic detection capability for corrupted data. This paper describes the benefits of adding the design parameter of increased data block size for modular addition checksums, combined with an empirical approach to modulus selection. A longer processing block size with the right modulus can provide significantly better fault detection performance with no change in the number of bytes used to store the check value. In particular, a large-block dual-sum approach provides Hamming Distance 3-class fault detection performance for many times the data word length capability of previously studied Fletcher and Adler checksums. Moduli of 253 and 65525 are identified as being particularly effective for general-purpose checksum use.**

## I. INTRODUCTION

Modular addition checksums are ubiquitous for protecting the integrity of data in communication and storage software. While more capable (and complex) error detection and correction codes have become common as computational speeds have increased, the humble checksum is still a mainstay in many application domains, and is likely to remain so for the indefinite future.

The main attraction of checksums is their simplicity, especially for low-resource embedded system applications. Simply adding up a series of data values in a sequence to get an integrity check value is about as simple as it gets. Even the dual-sum approaches (described in detail below) are still just a pair of running sums.

To be sure, more sophisticated coding approaches have their place. But that sophistication comes at the cost of algorithmic complexity and more demanding computational requirements. For example, Cyclic Redundancy Checks (CRCs) are very useful, and provide far superior fault detection capabilities to checksums [Koopman04], but are beyond the scope of this paper.

The search for improved checksums has extended across many decades, with the most notable innovation being the advent of the dual-sum approach by Fletcher circa 1982 [Fletcher82]. Other work has examined using a different modulus for the modular sum operation, such as the Adler checksum in the 1995 [Adler]. A more detailed discussion of previous work in checksums can be found in [Maxino09].

In this work we introduce a new parameter for checksum algorithms by considering larger block sizes for checksum processing. We also revisit the basis for modulus selection and find that an empirical approach reveals improved moduli compared to those typically used in existing implementations.

The contributions of this paper include: identifying better moduli to use in modular addition checksum algorithms, extending existing checksum algorithms to provide better performance via a large-block approach, showing that an large-block dual-sum checksum can provide Hamming Distance 3-class performance (i.e., detection of all two-bit faults for practical purposes) at multiples of data word sizes beyond current checksum algorithms, and provide an explanation for why large-block checksum approaches are so effective.

The remainder of this paper is organized as follows. Section II defines terminology and reviews commonly used checksum algorithms, including single-sum and dual-sum approaches. Section III reviews the experimental methodology for this empirically-driven exploration of improved checksum performance. Section IV describes improved modulus selection results. Section V presents larger block size results for single-sum checksums. Section VI presents the results of applying improved modulus selection and larger block sizes to dual-sum checksums. Section VII explains the fault detection mechanism involved with large block size checksums. Section VIII provides conclusions.

## II. PREVIOUS CHECKSUM ALGORITHMS

Checksum computations of interest for this paper take a collection of data (the data word), break that data into blocks of data, perform a modular addition across the data blocks within the data word to create a check value, and store that check value with the data word to create a code word. That code word can later be checked for integrity by recomputing the checksum from the data word and comparing that result to the stored check value. The following subsections describe this idea in more detail.

### A. Terminology

A *data word* is an ordered collection of data values for which integrity protection is sought via a checksum computation. For the purposes of this paper, a data word is a sequence of bytes of

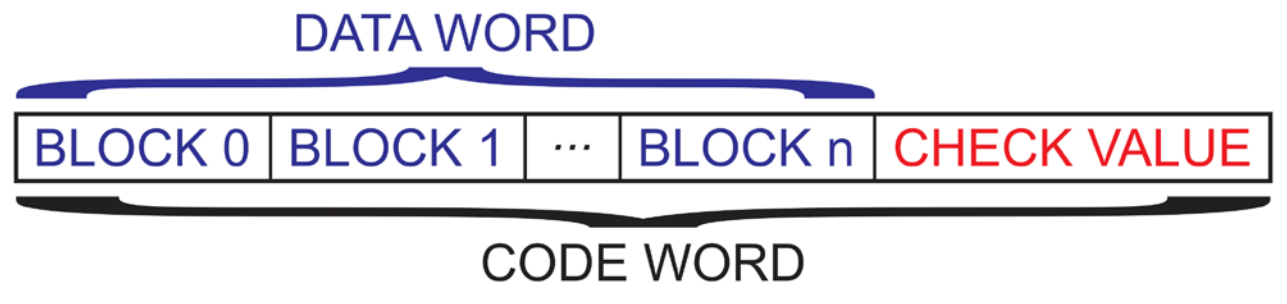

*Figure 1. Data organization of a code word.*

data, with the entire sequence of bytes considered a single data word. (Figure 1.)

A *check value* is the result of a checksum calculation. It might be based on a *single-sum algorithm* that performs a single running modular sum of all the bytes in the data word. It might instead be a *dual-sum algorithm* that performs a pair of coordinated running modular sums of all the bytes in the data word (more detail on this later). Data words can be of any length, whereas check values are typically comparatively smaller, often one, two, four, or eight bytes in size. For a single-sum algorithm, the check value is the same size as the running sum (e.g., a two-byte running sum gives a two-byte check value for a 16-bit single-sum checksum). For a dual-sum algorithm, each modular sum is half the size of the check value, with two modular sums concatenated to form a single check value (e.g., a pair of one-byte running sums are concatenated to give a two-byte check value for a 16-bit dual-sum checksum).

The check value is typically appended to the data word to form a *code word*. For the purposes of this paper, a code word is a sequence of bytes consisting of a data word followed by the bytes of the check value computed from that data word. This pair of data word plus check value is called a code word, in keeping with terminology from the area of error coding theory. The placement of the check value within the code word might be important in some circumstances, but does not affect the fault detection properties studied in this paper.

When computing a checksum, the data word is divided into a sequence of *blocks*. Those blocks are sequentially fed into a checksum algorithm to produce a check value. In the general case, there are multiple blocks in a data word. A policy is set for a computation involving a partial block that might be encountered at the end of a data word. We assume the policy is padding any missing bytes in the last data word with zeros. In the special case that the block size is larger than the data word, a single zero-padded block is processed.

A *checksum* algorithm computes a digest or hash of the blocks composing the data word to produce a check value. For the purpose of this paper, the computation is based on modular addition, in which each sum is modulo some specified *modulus value*, leaving a remainder after division of the sum by the modulus. (The use of the word "digest" in this paper does not connote any security properties. Checksum calculations are inherently insecure, and intended for use only in mitigating non-malicious data corruption.)

The modulus chosen must be in a range to produce the correct sum size, which means the modulus value should be between $2^{k-1}+1$ and $2^k$ inclusive for a k-bit sum. (For example, an 8-bit sum should have a modulus between 129 and 256, inclusive.) Not all moduli perform equally well, so modulus selection is an important design parameter for a checksum algorithm.

A code word is said to be *valid* if the check value in the code word matches a check value result computed from the data word portion of that same code word. An invalid code word is known to have been corrupted.

A valid code word could be uncorrupted. But a valid code word could also, by chance, have had its bits corrupted in a pattern that transforms it from the original valid code word to some other different, but serendipitously valid, code word. Such a serendipitous transformation between code words results in an *undetected fault*.

In keeping with previous work, the probability of an undetected fault $P_{ud}$ is the probability that any particular code word contains an undetectable (by the checksum algorithm) fault. This probability applies to all code words subject to potential faults rather than just the ones that would be known to be faulty by an omniscient observer. Note that in a typical case there will be many more detectable faults than undetectable faults, so system level interventions such as disregarding an obviously fault-prone communication channel or data storage device can be used to supplement checksums as part of a system fault management strategy [Koopman15].

*Bit Error Rate (BER)* is a typical parameter used in evaluating checksum performance. The assumption is that binary symmetric bit value inversions ("bit flips") will occur with a fixed, random independent probability across the length of the code word. Each bit in the code word is subject to an independent probability (the BER) of suffering an inversion in which a "0" bit is flipped to a "1" or a "1" bit is flipped to a "0". Note that these bit faults can occur to the entire code word, including the check value.

We use a BER of $10^{-6}$ for evaluation, meaning that each bit in a code word has a probability of 1 in 1,000,000 of being inverted. For the code word lengths we study (up to 32K bits plus the check value), the predominant fault modes will be single-bit faults (approximately 1 in 31.5 code words), two-bit faults (approximately 1 in 1923 code words), and three-bit faults (approximately 1 in 175,971 code words at that maximum code word length). Having more than 3% of messages corrupted is excessive for many real world data use situations, so this is a fairly pessimistic BER than emphasizes exposure to multi-bit faults more than a lower BER would. Changes to BER and code word length would affect the relative contribution of 1-, 2-, 3-, and other bit faults to the overall $P_{ud}$, but not the ability of a particular checksum algorithm to detect a given number of bit faults at a given code word length.

The important effectiveness metric for a checksum algorithm is its ability to minimize undetected faults, meaning that *lower* $P_{ud}$ means the checksum is *more effective*. $P_{ud}$ will naturally become lower with a lower BER and shorter code word lengths, because there are fewer corrupted codewords that tend to have fewer numbers of bit faults. So in that sense the effectiveness curves in this paper for long code words and fairly high BERs are pessimistic. But the $P_{ud}$ curves do, however, illustrate the relative performance effectiveness of different checksum algorithms in a relative sense even if the BER were to be different. In general, effectiveness differences will increase even further with lower BER as one- and two-bit faults become comparatively more frequent compared to larger numbers of bit faults.

A *Hamming Distance* (HD) for the purposes of this paper is the minimum number of bits in the code word that can be inverted to produce an undetected code word corruption. For the checksum algorithms discussed in this paper the HD is either two or three. A Hamming Distance of two (HD=2) means all single bit faults will be detected by the checksum, but at least one two-bit fault is undetectable due to conversion of the original code word to an incorrect, but valid, faulty code word. At HD=3, all single bit faults and all two-bit faults are detected, but at least one three-bit fault is undetected. For faults at and beyond the HD value, many faults are still detected, but some are undetected. (There are other potentially relevant fault detection properties, such as burst fault detection, that are beyond the scope of this paper.)

The HD is unaffected by the BER, but the number of multi-bit faults will increase as the BER increases given a fixed code word length. Thus, $P_{ud}$ will generally increase for higher BERs for any given checksum algorithm.

An important effectiveness consideration is that dual-sum algorithms offer HD=3 performance at comparatively short data word lengths, but will degrade to HD=2 performance at and beyond an algorithm-dependent data word length that we call the algorithm's *HD=3 capability*. HD=3 is highly desirable for a random independent bit error model due to the dramatically lower probability of a three-bit error compared to a two-bit error at any particular BER. Therefore, the longest possible HD=3 capability is especially desirable for general purpose checksum applications.

### B. Checksum Usage

A checksum algorithm is generally used for error detection in data transmission or data storage performing the following steps:

(1) The data word to be protected with an integrity check is placed into the code word, leaving room for a check value to be added in a later step. (Some communication systems initiate data transmission in parallel with computing the checksum. That difference does not matter for our purposes.)

(2) The check value is computed on successive blocks of data in the data word according to the checksum algorithm, resulting in a valid code word with check value $cv_0$. Some specified initial value is used to start the summing operation, which is assumed to be zero for this analysis.

(3) The bytes of check value $cv_0$ are placed into the remaining bytes of the code word, completing the code word.

(4) The entire code word is stored, transmitted, or otherwise sent into an environment in which it might suffer one or more corruptions in the form of bit inversions according to a BER-driven process. (Length changes and other fault models are relevant in the real world, but beyond the scope of the fault model used for this analysis.)

(5) An integrity check is performed by first using the checksum algorithm to compute a check value $cv_1$ from the contents of the potentially corrupted data word. Note that $cv_1$ might differ from $cv_0$ due to corruption of the data word – but the receiver of the code word has no way to know the ground truth of what $cv_0$ might have been, which is the motivation for performing subsequent steps in this procedure.

(6) The bytes in the check value field of the potentially corrupted code word are extracted from the code word bytes and assembled as different check value $cv_2$. Those bytes started holding a copy of $cv_0$, but $cv_2$ might not equal $cv_0$ if the check value field of the code word has been corrupted.

(7) The two recovered check values are compared: $cv_1$ (computed on the received data word), and $cv_2$ (recovered from the check value bytes in the code word).

(8) If $cv_1$ does not equal $cv_2$, the code word is invalid. Therefore, the code word has definitely been corrupted, even though there might not be enough information available to determine exactly which bits were corrupted.

(9) If $cv_1$ equals $cv_2$, one of two situations is true: either there has been no corruption of the codeword, or there has been a severe enough corruption (i.e., at least HD bits have been inverted) that the fault is undetectable by the checksum algorithm. In practice, the computation using the code word as a data source will accept the data as uncorrupted. But there will be a residual probability of an undetected corruption, $P_{ud}$, that could lead to an eventual system failure. The lower $P_{ud}$, the more effective the checksum algorithm.

The above steps apply to all checksum algorithms discussed in this paper. The differences among algorithms discussed have to do with the whether the algorithm is single-sum or dual-sum, and the algorithmic parameters of block size and modulus.

### C. Single-Sum Checksums

Classical checksum algorithms involve computing a single modular sum of block values drawn in sequence from the entire length of the data word. A generic description of such an algorithm is shown in Algorithm 1.

```
Initialize Sum_initial = 0

Iterate across each block i in data word:
   Sum_new = ( Sum_old + Block_i ) mod M

Check Value is the final Sum_new
```

*Algorithm 1: Single-sum checksum.*

In Algorithm 1, M is a selected algorithm-dependent modulus. At each iteration, Sum is updated with the next block from the data word with a single modular addition. When all blocks have been processed, the final value of Sum is used as the check value for the code word.

For example, a 16-bit check value with two-byte blocks (block size the same as the check value size) would process the data word in blocks of two bytes at a time. A 256-byte data word would therefore be processed as 128 two-byte blocks, yielding a two-byte modular addition summed check value. Typical single-sum checksum parameters are below:

- Twos8: (Two's complement addition)
  block size = 1 byte
  check value = 1 bytes
  modulus = 256
- Ones8: (One's complement addition)
  block size = 1 bytes
  check value = 1 bytes
  modulus = 255
- Prime8: (Largest prime modulus)
  block size = 1 bytes
  check value = 1 bytes
  modulus = 251
- Twos16:
  block size = 2 byte
  check value = 2 bytes
  modulus = 65536
- Ones16:
  block size = 2 bytes
  check value = 2 bytes
  modulus = 65535
- Prime16: (Largest prime modulus)
  block size = 2 bytes
  check value = 2 bytes
  modulus = 65521

The pattern is that a two's complement checksum uses a modulus of $2^k$ for a k-bit check value. A one's complement checksum uses a modulus of $2^k$-1 for a k-bit check value. A prime modulus (an abbreviation of "largest prime") uses the largest prime number less than $2^k$ for a k-bit check value. 32-bit variants are possible with a block size of 4 bytes, check value of 4 bytes, and a modulus chosen according to the check value size.

For performance purposes, it is helpful to note that two's complement checksums can simply use an 8-, 16-, or 32-bit register for the addition and ignore carry-outs (assuming use of ubiquitous two's complement CPU hardware). In that sense, all additions on finite-size binary integers are modular division with a modulus of $2^k$ for a k-bit hardware register – even if the modulo operation does not require an explicit division computation to be performed.

One's complement checksums can be implemented in practice by incrementing the running sum if a carry-out of the addition is detected. A lossless sum of two k-bit numbers in principle requires k+1 bits to store, such as an 8-bit sum of example values 250+10=260, which requires 9 bits instead of 8 to represent. One's complement addition increments the k-bit sum if that top-most k+1$^{st}$ bit would have been needed to represent the sum (e.g., the carry-out of an 8-bit addition operation), for this example resulting in an 8-bit sum of ((250+10) mod256 + 1) = 5. This wrapping of the carry-out bit makes one's complement checksums less vulnerable to bit faults on the top-most bit of a block [Maxino09].

Prime checksums use the largest prime number that fits in a block-size number of bits as the modulus. The general idea is that a prime number typically has a mix of zero and one bits in its representation, promoting mixing among bits in the sum via the modulo operation to create a check value that is more effective. Smaller prime numbers might be used instead, but the thinking is that the largest prime makes the most efficient use of the available check value range. For example, a prime modulus of 251 supports a check value range of [0..250] whereas a prime modulus of 239 supports a smaller check value range of [0..238], which all things being equal would give better odds of detecting faults. (As it turns out, all things are not equal for small numbers of bit faults that dominate effectiveness under a BER fault model.)

*D. Dual-Sum Checksums*

A more advanced class of checksums was introduced by Fletcher's work [Fletcher82]. In the Fletcher approach, a pair of running modular sums is used instead of a single sum. The first sum, which we denote SumA, is a conventional modular checksum that accumulates a running modular sum of all blocks in the data word.

The second sum in Fletcher's algorithm, SumB, is a running sum that is updated not by summing block values, but rather by summing the old version of SumB with the new version of SumA for each block being processed. The check value result is the concatenation of SumA and SumB.

> Initialize $SumA_{initial} = 0$; $SumB_{initial} = 0$
>
> Iterate across each block *i* in data word:
> $SumA_{new} = (SumA_{old} + Block_i) \bmod M$
> $SumB_{new} = (SumB_{old} + SumA_{new}) \bmod M$
>
> Check Value is final $SumA_{new}$ concatenated with $SumB_{new}$

*Algorithm 2: Dual-sum checksum.*

As with Algorithm 1, for Algorithm 2 M is the modulus, with the same modulus being used for both sums. All blocks from the data word are processed in a running sum approach, with the pair of sums updated as each block is processed.

The concatenation operation means that the size of the check value is twice the size of each sum (e.g., two 16-bit sums are paired to produce a 32-bit check value). The sizing notation for the checksum is the check value size, so two 16-bit sums would be designated as a 32-bit dual-sum checksum.

The two well-known existing dual-sum approaches are the Fletcher checksum [Fletcher82] and the Adler checksum [Adler]. The Fletcher checksum uses a one's complement modulus, and the Adler checksum uses a largest-prime modulus. (Some implementations of the Fletcher checksum are said to use a two's complement modulus, but we disregard them to avoid confusion. They have uniformly worse performance than the proper one's complement implementation.)

- Fletcher-16:
  block size = 1 byte
  check value = 2 bytes
  modulus = 255
- Adler-16:
  block size = 1 byte
  check value = 2 bytes
  modulus = 251
- Fletcher-32:
  block size = 2 bytes
  check value = 4 bytes
  modulus = 65535
- Adler-32:
  block size = 2 byte
  check value = 4 bytes
  modulus = 65521

As with single-sum checksums, dual-sum checksums can be scaled up or down in size by selecting appropriate parameters. Key properties of these checksums are that the check value is twice the block size, and the modulus is chosen to fit within the block size so that there is room for all the bits in both SumA and SumB in the check value.

As discussed in Fletcher's original paper [Fletcher82], dual-sum checksums have the property that they are inherently HD=3 through the number of data word bytes equal to the modulus minus one, which in the case of Fletcher16 is 255-1=254 bytes. This is because the SumB is effectively multiplying each SumA value by its position in the data word. For example, for a 255-byte data word the value of Sum B is:

$$255*SumA_0 + 254*SumA_1 + 253*SumA_2 + \ldots + 1*SumA_{254} \quad (1)$$

Because the SumB addition is modulo 255, the contribution from $255*SumA_0$ is zero, causing it to have no effect on SumB. Contributions from all blocks before the most recent 255 blocks to be lost from SumB (we call this the *rollover length* for a dual-sum checksum computation). That leaves the code word vulnerable to two-bit faults in the data word exactly 255 bytes apart that cancel each other out, since at that point SumA is all that is protecting against that type of fault, and is vulnerable two such faults the same as an 8-bit single-sum checksum would be.

All dual-sum checksums have a rollover length equal to the value of the modulus due to this mechanism. Nonetheless, having a guarantee of HD=3 up to a length of Modulus-1 is valuable, and this is something that we shall enhance with a large-block approach later in this paper.

*E. Baseline Checksum Performance*

Maxino and Koopman [Maxino09] previously evaluated the performance of checksum algorithms. Figure 2 shows simulation results from the study reported in this current paper in a format to facilitate comparison with fig. 6 of that previous publication, using a BER of $10^{-5}$. (That BER was suitable for the shorter messages previously studied.) That previous work found that one's and two's complement single-sum checksums differed in vulnerabilities to two-bit faults in the topmost block position for two's complement checksums, but otherwise had the same performance. The difference is difficult to discern on a semi-log plot.

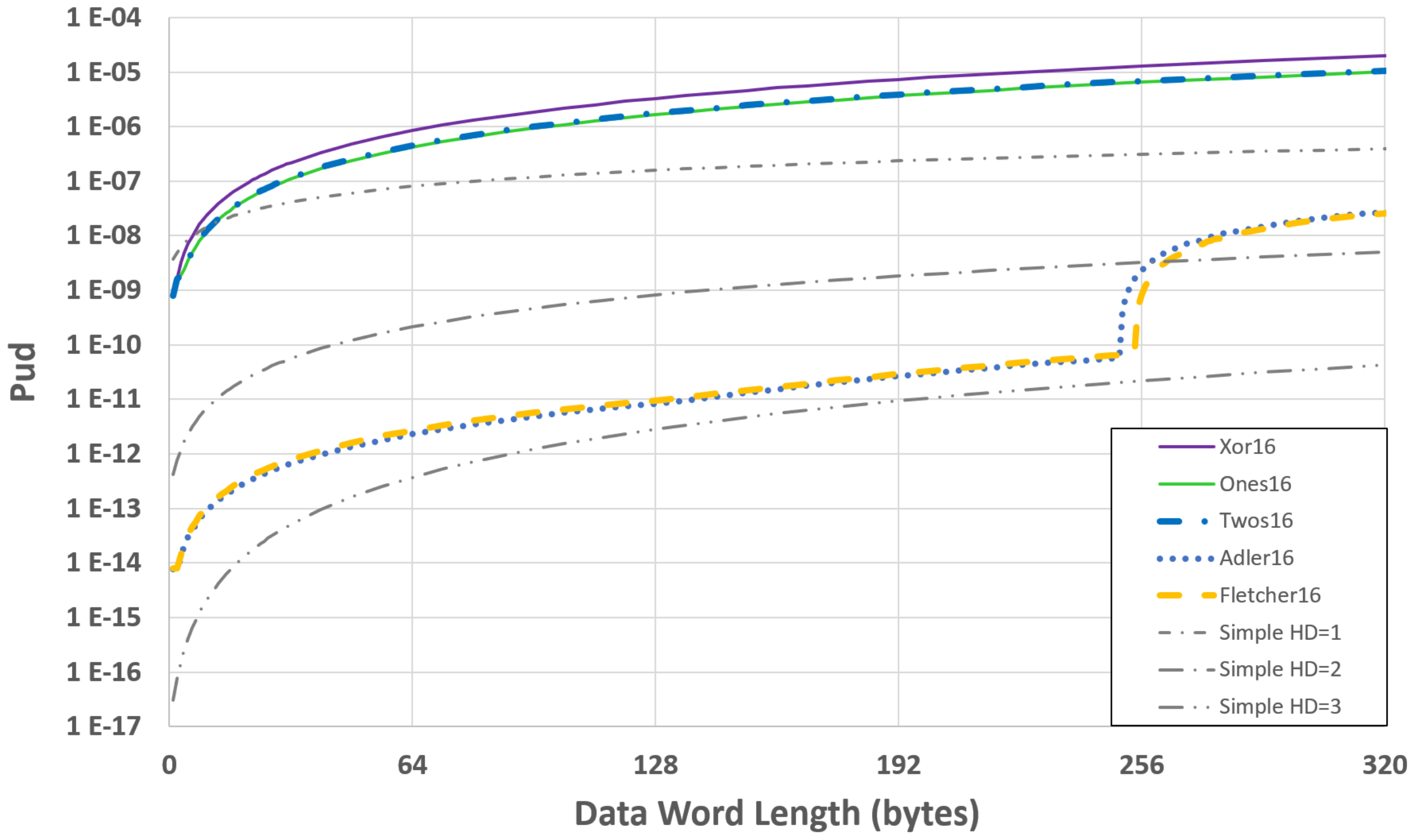

*Figure 2. Legacy checksum performance at BER of $10^{-5}$. Note that this is different than the BER of $10^{-6}$ used in other figures.*

That previous work also found that Fletcher and Adler checksums had roughly comparable effectiveness. Both achieved HD=3 up to one byte data word lengths shorter than the modulus value, then transitioned to effectiveness dominated by 2-bit faults at longer data word lengths. This corresponds to the behavior predicted by the analysis of their rollover behavior.

Our figure 2 adds three lines for idealized checksum performance. The "Idealized HD=1" assumes a fictitious 16-bit checksum algorithm that detects all faults with a probability of precisely 1/65536, which assumes completely uniform distribution of check values across the space of all possible check values. In other words, every faulty code word is assumed to be detected with probability of 1/65536 regardless of how many bit faults it has. One might expect this sort of performance with, for example, a 16-bit hash value generated from a cryptographically secure hash function.

"Idealized HD=2" assumes all one-bit faults are detected, but there is a uniform distribution of check values for all other faults with a probability of undetected faults of 1/65536. (This idealized HD=2 curve corresponds to the curve denoted "1/2^k" in [Maxino09].) "Idealized HD=3" assumes all one- and two-bit faults are detected, again with a probability of undetected faults of 1/65536. The point of these lines is to provide a reference for idealized fault detection effectiveness, and not to imply that such checksums might actually be implemented in a simple and efficient way.

In figure 2 we can see that single sum checksums do worse due to the poor mixing of bits via an addition function for all but the shortest data word lengths. Dual-sum checksums (Fletcher and Adler) do somewhat worse than idealized HD=3 effectiveness up to their HD=3 capability, then operate above the idealized HD=2 curve.

The findings in this paper have reproduced the previous checksum findings from [Maxino09] from scratch using a different simulation approach and all-new code base, validating both that previous work and helping to validate the code base used for this newer work. [Maxino09] contains a much more extensive treatment of previous work and analysis of existing checksums which is not repeated here in the interest of space.

## III. Methodology

Figure 2 and other fault detection performance results were created with a purpose-built Monte Carlo simulation framework written in the C programming language. Simulations were run as single-threaded applications batched across 24 physical processor cores. The framework operates on the following principles, staying within the context of the overall steps in the use of checksums from section IIB described above:

(1) A 32-bit PCG generator is used to generate the random byte stream [PCG]. The PCG generator is seeded differently for each run, based on time of day. (This suffices to produce dramatically diverging simulation runs. Cryptographic security of the pseudo-random number stream is not of concern for this purpose, so time of day is as good a practical source of different seeds as any.) Regression testing of the framework is done using a constant initialization seed value for repeatability.

(2) For each experimental simulation step (an *experiment* in our terminology), a data word of a specified length is created using pseudo-randomly generated data bytes.

(3) The check value using the checksum algorithm being investigated is computed to create a code word.

(4) A pseudo-randomly selected bit is inverted within the code word.

(5) The checksum algorithm is run to determine if the known-corrupted codeword is valid. If it is valid, it must be an undetected fault because the codeword has been explicitly subjected to a known fault injection, so the fault counter for that number of bit inversions is incremented.

(6) Steps (4)-(5) are repeated for additional, increasing numbers of bit faults with that same code word. At least one- through five-bit faults are evaluated for each experiment.

(7) New pseudo-randomly generated data words are created and tested via fault injection, repeating steps (2)-(6) to run a set of many experiments. A set of experiments (typically tens to hundreds of millions of experiments at a single data word length) results in a single *data point* of undetected faults for a range of number of inverted bits at a specific data word length for a specific checksum algorithm.

(8) A spreadsheet is used to accumulate multiple data points and convert the ratio of undetected faults to number of accumulated experiments into a $P_{ud}$, taking into account data word length and BER.

(9) A curve is plotted based on the total number of undetected faults across the collected data points. If the curve is not reasonable smooth, additional data points are collected using the above procedural steps.

The data points are not a direct $P_{ud}$ simulation result, but rather a tuple of undetected tallies for a number of experiments with different fixed numbers of pseudo-randomly injected bit faults at a specific data word length for a specific checksum algorithm. This produces results that can yield better insights than a simulation based primarily on a probability-based bit inversion strategy, because it identifies the contribution of different numbers of bit errors to the checksum performance. For example, the HD=3 capability can be determined explicitly by looking for the shortest length with even a single non-zero undetected two-bit fault in simulation results. That permits increased understanding and confidence rather than having to try to infer the inflection point from comparatively small changes in a direct simulation of $P_{ud}$ via random fault injection using the per-bit BER probability.

$$Pud = Pud_{(HD)} + Pud_{(HD+1)}$$

$$Pud = \text{combin}(c, HD) * \text{UndetectedFraction}_{HD} * BER^{HD} * (1 - BER)^{(c - HD)} + \text{combin}(c, HD + 1) * \text{UndetectedFraction}_{HD+1} * (BER^{(HD+1)}) * (1 - BER)^{(c - HD + 1)}$$

*Figure 3. Approximate $P_{ud}$ calculation from [Koopman15].*

Pud is computed via a spreadsheet that uses Equations (1) and (2) of [Koopman15] (see figure 3), extended to encompass the available undetected bit fault information (one-bit to at least five-bit faults). Substantive contributions to $P_{ud}$ are for practical purposes only made by up to the first two or three non-zero undetected fault weights.

The spacing of data points along the data word is adjusted for each checksum algorithm to improve fidelity when there are large changes in curvature. At short data word lengths and near the HD=3 capability for dual-sum checksums, data points are taken for every consecutive byte length.

The number of experiments in a data point varies depending on the checksum algorithm. Lower $P_{ud}$ results demand more experiments to collect enough examples of undetected faults to create smooth $P_{ud}$ curves. Regardless of the algorithm, however, the number of experiments per data point ranges from the tens of millions to several billion, and each curve has in excess of 100 data points at various data word lengths.

In terms of raw experimental results, it is desirable to have at least several hundred undetected faults for the first non-zero number of undetected bit faults to achieve smooth $P_{ud}$ curves. The number of undetected 2-bit faults is quite small at the HD=3 capability, so more experiments are run in that vicinity to ensure that 2-bit faults are consistently identified at all data points above the HD=3 capability (and no surprise 2-bit faults are undetected just below that data word length).

In practice, achieving a smooth $P_{ud}$ line is a more demanding measure of statistical significance than more typical experimental evaluation approaches. This happens because each data point on the line is computed independently, but the data points are relatively dense on a line. This results in a visual measure of fluctuation from point to point that appears smooth if the fluctuations of data point values due to stochastic noise is less than about 1-2% on the semi-log scales being used. (To be clear, all plots shown in this paper are drawn with point-to-point line segments and are not fitted curves. Under these conditions, visual analysis of curve smoothness is a surprisingly sensitive technique.)

An additional observation is that even very bumpy curves closely approximate final curves obtained via accumulation of data points. In other words, as data points accumulate the curve smooths out rather than changing its basic shape. Visual observation of decline in curve bumpiness turns out to be an excellent measure of simulation progress as additional data points are collected from experimental batches.

Limitations of available computer time dictate that experiments be run until the plot is smooth without re-running in evenly-sized distinct data sets, so statistical measures beyond data plot smoothness were impractical. That having been said, curve smoothness and the dramatic differences in effectiveness between algorithms, confirmed by analysis, make it clear that the results in this paper indicate real effects and not statistically ambiguous findings.

The remaining effectiveness plots use a BER of $10^{-6}$ as described previously. For consistency, figure 4 shows legacy checksum effectiveness at this BER.

## IV. Alternate Moduli

Our results show that a more empirically-based selection of modulus for the modular addition operation can improve fault detection effectiveness. As noted previously, moduli are traditionally selected based on corresponding to two's complement or one's complement addition, or a prime modulus. However, other alternatives not dependent upon the factorization of the modulus can provide better performance.

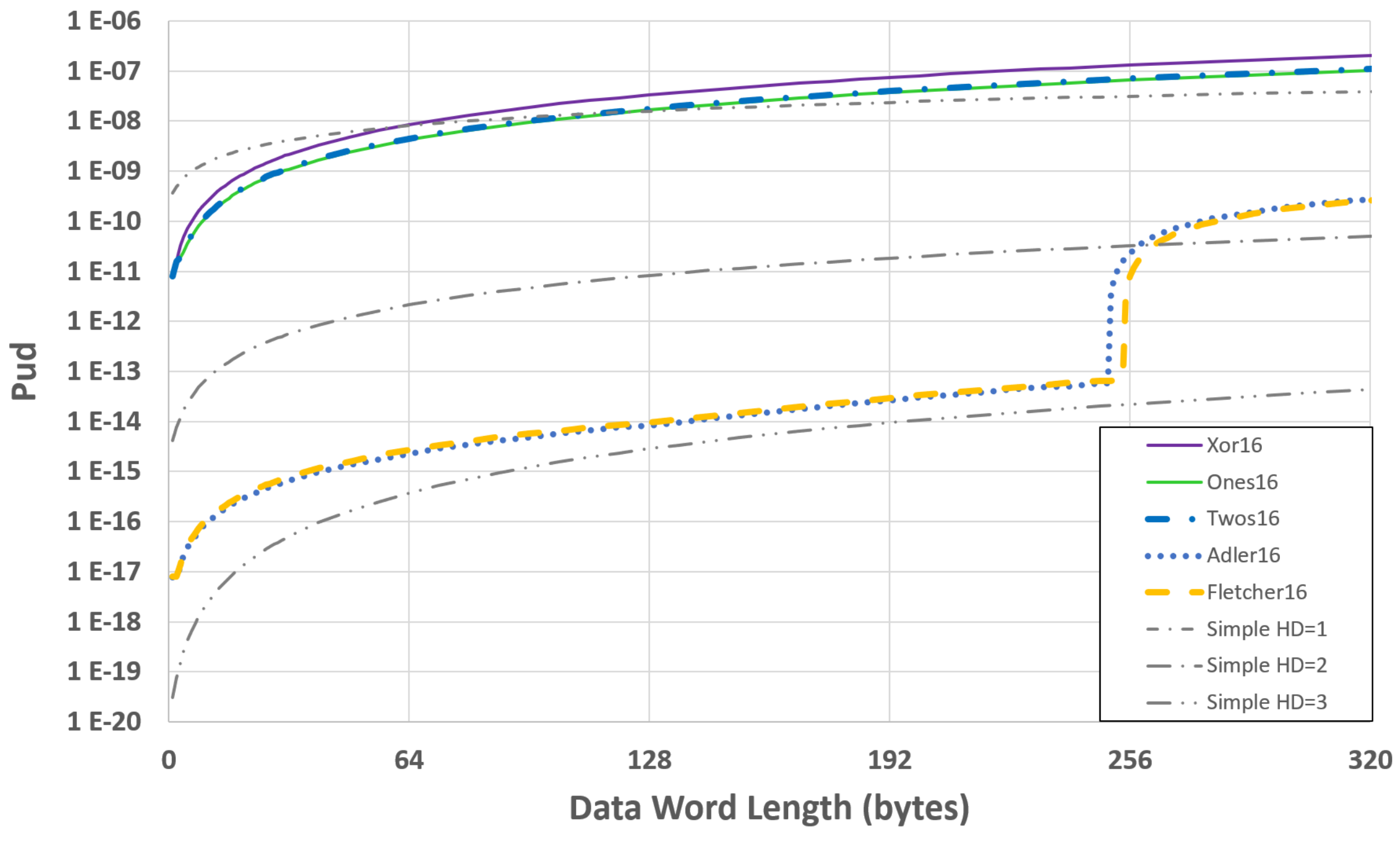

*Figure 4. Curves from figure 1 replotted with BER=$10^{-6}$.*

*A. Modulus Selection Criteria*

As previously discussed, a two's complement modular addition simply ignores any carry-out bits larger than the integer size being used. A one's complement is in principle a mod 255 operation, but in practice amounts to "wrapping" the carry-out bit out and back into the sum as an additional post-sum increment of the result if the carry-out bit was a "1".

The only mixing promoted by either a two's or one's complement addition checksum is the normal inter-bit-position carries inherent to addition. The effect of that bit mixing is indeed helpful, and can be seen in the performance difference between an XOR checksum, which has no inter-bit carries, compared to a two's complement addition. However, even with inter-bit carries, sums are still vulnerable to two-bit faults that occur in the same bit position for block values that do not happen to involve a carry operation affecting those bit positions.

A way to improve mixing among non-adjacent bit positions is to use a different modulus that involves a more substantive division operation, mixing bits beyond the addition carry effects. This can be done by selecting a modulus that has a mix of 0 and 1 bit values in its representation while yielding an output value that provides the same number of bits as the running sum.

An intuitive candidate for an alternate modulus is using a prime number. A typical justification is that a prime number has no common factors with two's complement modulus, since all two's complement moduli are even powers of two.

A further tradeoff is that a larger modulus makes better use of the available representation space of the check value as discussed earlier. However, the largest odd modulus (255 for 8-bit checksums) turns out to be a poor choice, so additional care is required in modulus selection.

*B. Experimental Modulus Selection*

Simulations for 8-bit and 16-bit single-sum checksums showed that even moduli values were a uniformly poor choice. Odd moduli were much better, with the choice of odd modulus value making little difference, with a few exceptions.

For 8-bit moduli, the values 201 and 129 are significantly worse than other moduli, even ones very close to them, for no readily discernable reason. The modulus 255 is slightly worse than other moduli for small, odd numbers of bit faults, but about the same for even numbers of bit faults. The largest prime modulus 251 is slightly better than 255, but not dramatically so, and not substantively better than other odd moduli. Smaller prime numbers used as moduli similarly show no substantive performance difference (227, 229, 233, 239, and 241). Table 1 shows representative effectiveness, with prime moduli bolded.

For 16-bit moduli there are similarly some poor candidates, but they are few. Prime moduli of either size do not have any distinct advantage. They perform about the same as most other odd moduli.

We will discuss in the next section that there are compelling reasons to select specific moduli for large-block checksums. For traditional single-sum checksums a largest-prime modulus is as good a choice as most, but not distinctly better than many non-prime moduli. A one's complement modulus loses effectiveness slightly for odd numbers of bit faults, but if it can be implemented without using a division instruction, modulus 255 might still be advantageous due to improved computational speed.

*Table 1. Undetected faults for single-sum checksum with varied moduli. 90 million data points per modulus. 128 byte data word; 1 byte check value; 1 byte block.*

| | Fraction of undetected faults | | | | |
|---|---|---|---|---|---|
| Modulus | 1-bit | 2-bit | 3-bit | 4-bit | 5-bit |
| 255 | 0% | 6.21% | 1.17% | 1.38% | 0.72% |
| 253 | 0% | 6.20% | 1.02% | 1.38% | 0.66% |
| **251** | **0%** | **6.21%** | **1.02%** | **1.34%** | **0.66%** |
| 249 | 0% | 6.20% | 1.02% | 1.34% | 0.66% |
| 247 | 0% | 6.21% | 1.02% | 1.34% | 0.65% |
| 245 | 0% | 6.20% | 1.02% | 1.30% | 0.64% |
| 243 | 0% | 6.21% | 1.02% | 1.31% | 0.64% |
| **241** | **0%** | **6.20%** | **1.02%** | **1.34%** | **0.67%** |
| **239** | **0%** | **6.20%** | **1.02%** | **1.34%** | **0.67%** |
| 237 | 0% | 6.20% | 1.02% | 1.30% | 0.63% |
| 235 | 0% | 6.20% | 1.02% | 1.31% | 0.62% |
| **233** | **0%** | **6.20%** | **1.02%** | **1.30%** | **0.65%** |
| 231 | 0% | 6.20% | 1.02% | 1.30% | 0.65% |
| **229** | **0%** | **6.20%** | **1.02%** | **1.30%** | **0.64%** |
| **227** | **0%** | **6.20%** | **1.02%** | **1.30%** | **0.67%** |

Modulus selection tradeoffs change dramatically for large-block checksums. The moduli recommended for large-block checksums will be comparable in performance to prime modulus checksums for small blocks, but have added flexibility for large-block application. For example, we shall see that 253 is a better modulus choice than either 251 or 255 for large block checksums.

## V. Single-Sum Large-Block Checksum Processing

Beyond the modulus, another parameter that can be varied in defining a checksum algorithm is the block size. Modular addition can do more than gracefully handle addition overflow bits. It can also perform a range reduction operation on a block size much larger than the sum. This does not redefine the checksum algorithmic description, but does re-envision what the operations in that algorithm are doing.

*A. Large-Block Modular Addition*

As stated earlier, the heart of a checksum operation is the modular sum:

$$Sum_{new} = ( Sum_{old} + Block_i ) \bmod M \tag{2}$$

The presumption in previous checksum operations is that each block and the running sum are the same size. But what if the block is significantly larger, such as a one-byte check value with a 4-byte or even 8-byte block size? It turns out this can dramatically improve error detection effectiveness.

The modular checksum operation can be rewritten, exploiting the commutativity of modular addition, to:

$$Sum_{new} = (\ Sum_{old} + \ (Block_i \bmod M)\ \ ) \bmod M \qquad (3)$$

Adding the additional mod M operation to $Block_i$ does not change the mathematical result. But it does call attention to the fact that the mod operation in modular checksum is actually performing two functions concurrently if block size is increased:

(1) Range reduction of the block to the integer size of the running sum (the inner "mod M" reduces the block to be the same number of bits as the running sum, assuming the modulus is sized to do this).

(2) Wrapping any overflow from the addition beyond the modulus size back into the running sum (the outer "mod M").

The range reduction step (1) can become a potent bit mixing operation using a natively supported division instruction. This is because remainder after division is a convolution operation, resulting in bits of the divisor in essence being used to stir the bits of the dividend. The larger the block size, the more effective this bit mixing will be.

Increasing the block size provides improved effectiveness for single-sum checksums, and dramatic improvements for dual-sum checksums.

## B. *Large-Block 8-Bit Single Sum*

A *large-block* checksum has a block size larger in terms of number of bits than the size of the addition being used in the checksum addition. More precisely, it has a block size larger than the next power of two larger than the modulus.

As a concrete example, a single-sum checksum with a check value of 1 byte traditionally has a block size of 1 byte. In a large-block checksum, the block size might be 2 bytes, 3 bytes, 4 bytes, or larger. A block size of 8 bytes would be readily supported on a processor with 64-bit architected data registers, and support is routinely available for 128-bit values on some computing platforms. Similarly, a large-block dual-sum checksum processes the data work in blocks larger than the size of each individual sum.

This paper explores the effects of blocks in size up to 16-byte blocks (128 bits), which is the largest convenient integer size on commonly available current computers. We expect 4-byte and 8-byte blocks to be especially common in practical implementations, so we present results for an assortment of block sizes. From a programming point of view, this is done by using large integer variables for computing the checksum, processing a set of multiple bytes at a time from the data word. The algorithms are the same as previously described Algorithms 1 and 2. It is simply that the blocks are larger. Multiple bytes from the data word are processed for each sum, and some care must be exercised to avoid integer overflow on intermediate sum results if they completely fill the bits of a declared variable of a given size.

Modulus selection for large-block modular sums is much more critical than for normal-block modular sums. Some moduli have a dramatic reduction in performance at larger block sizes, while others do not. Perhaps surprisingly, a prime modulus is not necessarily the best choice.

*Table 2. Undetected faults for single-sum checksum with varied moduli with large block size. 128 byte data word; 1 byte check value; 8 byte block.*

| | Fraction of undetected faults | | | | |
|---|---|---|---|---|---|
| Modulus | 1-bit | 2-bit | 3-bit | 4-bit | 5-bit |
| 255 | 0% | 6.20% | 1.17% | 1.38% | 0.72% |
| 253 | 0% | 0.73% | 0.39% | 0.40% | 0.39% |
| **251** | **0%** | **2.03%** | **0.37%** | **0.45%** | **0.40%** |
| 249 | 0% | 1.30% | 0.33% | 0.46% | 0.38% |
| 247 | 0% | 1.42% | 0.42% | 0.42% | 0.41% |
| 245 | 0% | 0.73% | 0.37% | 0.42% | 0.41% |
| 243 | 0% | 0.73% | 0.31% | 0.46% | 0.38% |
| **241** | **0%** | **4.16%** | **0.87%** | **0.84%** | **0.53%** |
| **239** | **0%** | **0.73%** | **0.41%** | **0.42%** | **0.42%** |
| 237 | 0% | 0.73% | 0.32% | 0.47% | 0.40% |
| 235 | 0% | 0.73% | 0.40% | 0.43% | 0.42% |
| **233** | **0%** | **1.73%** | **0.27%** | **0.47%** | **0.42%** |
| 231 | 0% | 1.66% | 0.42% | 0.56% | 0.42% |
| **229** | **0%** | **1.37%** | **0.46%** | **0.45%** | **0.44%** |
| **227** | **0%** | **0.73%** | **0.44%** | **0.44%** | **0.44%** |

Table 2 shows modulus performance for single-sum addition checksum at a data word length of 128 bytes and block size of 8 bytes.

From table 2 we can see that, at the longer block length of 8, fault detection can vary dramatically compared to block size 1 effectiveness shown in table 1. We consider undetected 2-bit faults since those will dominate the $P_{ud}$ results. Modulus 255 has effectiveness at a block size of 8 that is essentially unchanged compared to a block size of 1 (modulus 255 @ 6.20%). The largest prime modulus has somewhat better performance (modulus 251 @ 2.03%). But modulus 253, which has not previously been considered as an attractive checksum modulus candidate, has dramatically better performance (modulus 253 @ 0.73%, more than a factor of 8 improvement compared to block size 1 performance).

Visualizing the performance of different moduli at increasing block sizes reveals an interesting pattern.

Figure 5 shows the results of simulating single-sum checksum effectiveness for different moduli at a data word size of 128 bytes. (Other data word sizes sampled have substantially similar results.) The top line for 1-byte blocks shows essentially the same effectiveness on two-bit faults for all moduli. Subsequently lower charted lines show the fraction of undetected faults decreasing for moduli other than 255 – up to a point. At some modulus-dependent block size, effectiveness improvement stalls, increasing only marginally with increasing block size past that point.

Modulus 255 does not improve past block size 1. Modulus 251 gets stuck at block size of 3, yielding only marginal improvements past that point. On the other hand, modulus 253 and a few others keep providing improved performance up to a block size of 8 (and, as we shall see in a later figure, even longer than that).

From this graph, it is clear that modulus 253 is a better choice than 251 (the largest prime) for moderate to large block sizes. But the question is, why? The answer to why some moduli do better than others is related to the two-bit fault sensitivity of modular sums.

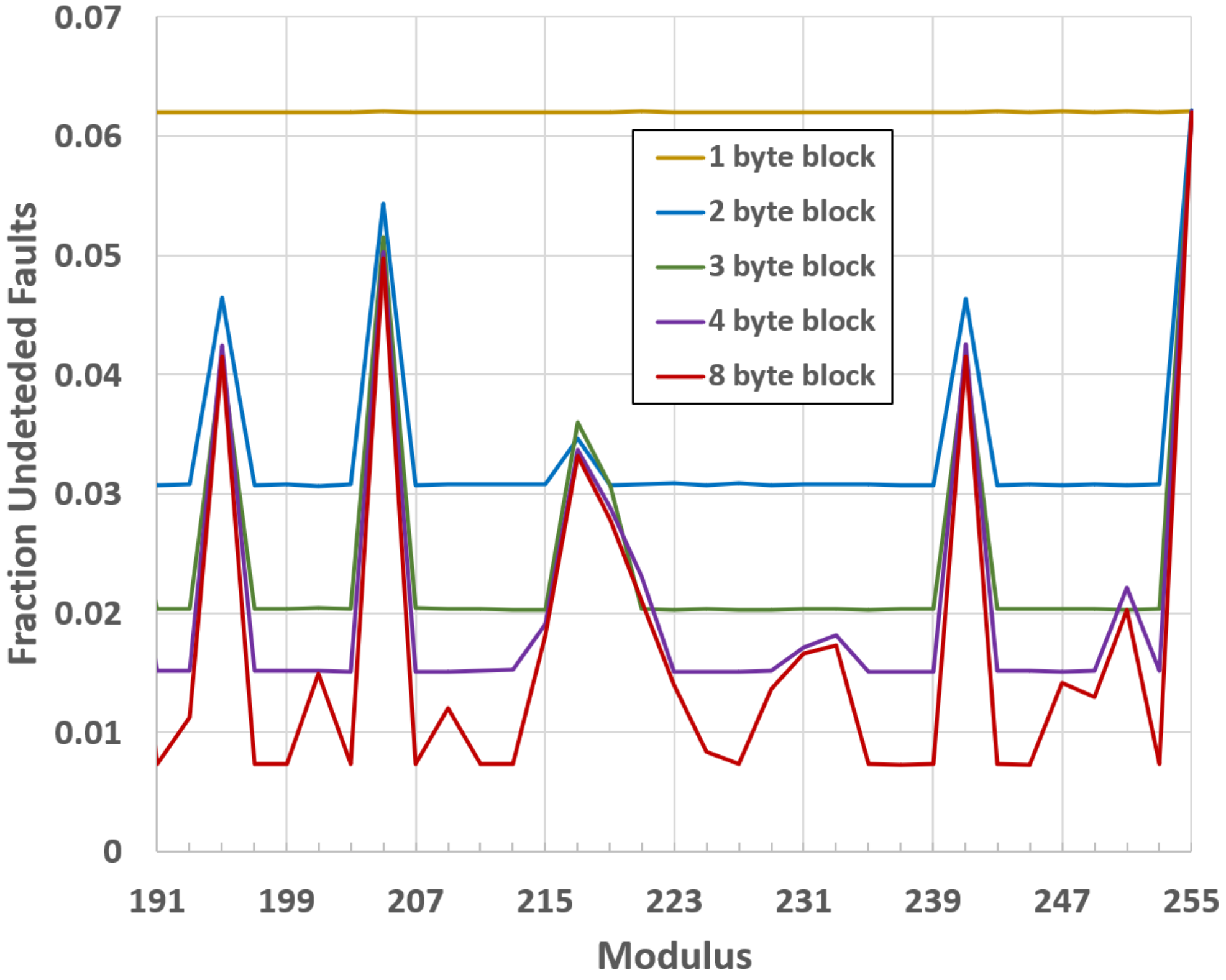

*Figure 5. Undetected fault fraction for odd moduli with increasing block size. Two-bit faults; 128-byte data words.*

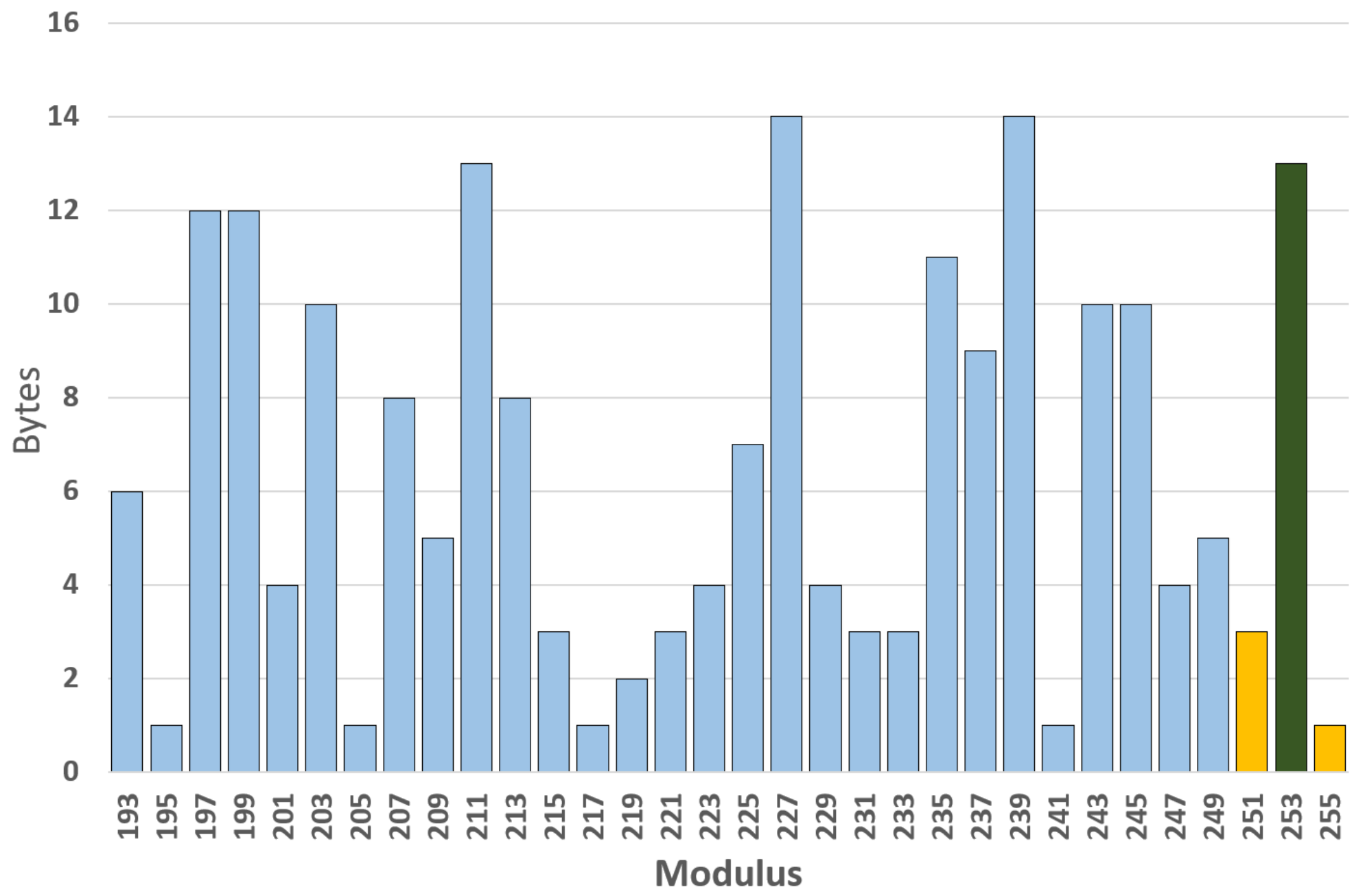

*Figure 6. Largest block size in bytes with no undetected two-bit faults from single-block modular division. (Considers faults in the data block and not check value faults.)*

### C. *Two-Bit Fault Sensitivity of Modular Sums*

There are two ways that a two-bit fault can be undetected in a single-sum checksum. The first mechanism is two bit faults in the same position in two data blocks. Those two faults can compensate for each other and result in the same checksum value. As an 8-bit example:

Uncorrupted: Data word: 0x00 00 Check Value: 0x00

Corrupted: Data word: 0x01 01 Check Value: 0x00

There is a different but similar undetected fault mechanism in which one fault occurs to a data block, and a second fault occurs in the same bit position of the check value. As an 8-bit example:

Uncorrupted: Data word: 0x00 00 Check Value: 0x00

Corrupted: Data word: 0x01 00 Check Value: 0x01

Again, the faults result in a valid code word and undetected two-bit fault. These fault mechanisms apply to all single-sum checksums.

Increasing the block size introduces a third possible mechanism to create undetectable two-bit faults. If two bits in the same block are inverted, it might be the case that the result of the "mod M" operation has the same value with and without the two-bit fault injection. If there is no difference in the output of the modulus range reduction into the running sum, there is no way for the checksum to detect that such a fault has occurred.

This means faults will be undetectable for block sizes larger than the running sum size if they satisfy this equality:

$$\text{Block mod M} = (\text{Block } xor \text{ Fault}) \text{ mod M} \quad (4)$$

where in the case of interest, the Fault value has exactly two "1" bits (i.e., exactly two bits are inverted in the Block value by the xor operation).

A different Monte Carlo simulation program was created to inject faults in pseudo-randomly generated integers from 1 to 16 bytes in size and determine if the remainder after division was the same. (This simulation did not compute a checksum value and did not attempt to corrupt bits in any check value – it was solely to look for pairs of bits that would be undetectable if corrupted as an input to a mod M operation for a particular modulus.)

In figure 6 each bar represents the maximum number of bytes in a block that could be used while avoiding two-bit faults within that same block that cancel each other out as just described. (To be sure, single faults in the block could result in changes to the remainder used in the sum. The only case of concern for this graph is one in which exactly two bit faults in the block result in an unchanged remainder value after applying the modulus.)

The height of the bars in figure 6 explains the patterns in figure 5 in which different moduli stopped performing well beyond a particular block length. Two bit faults via the first and second mechanism described above will still result in an undetected fault rate at all block lengths, driven by the ability of faults in the block to generate a one-bit result after the modulus is applied. But once the modulus being used is vulnerable to undetectable two-bit faults within the same block, that permits those faults to escape without the summing operation ever getting a chance to detect them, resulting in very little further effectiveness improvement with increased block size.

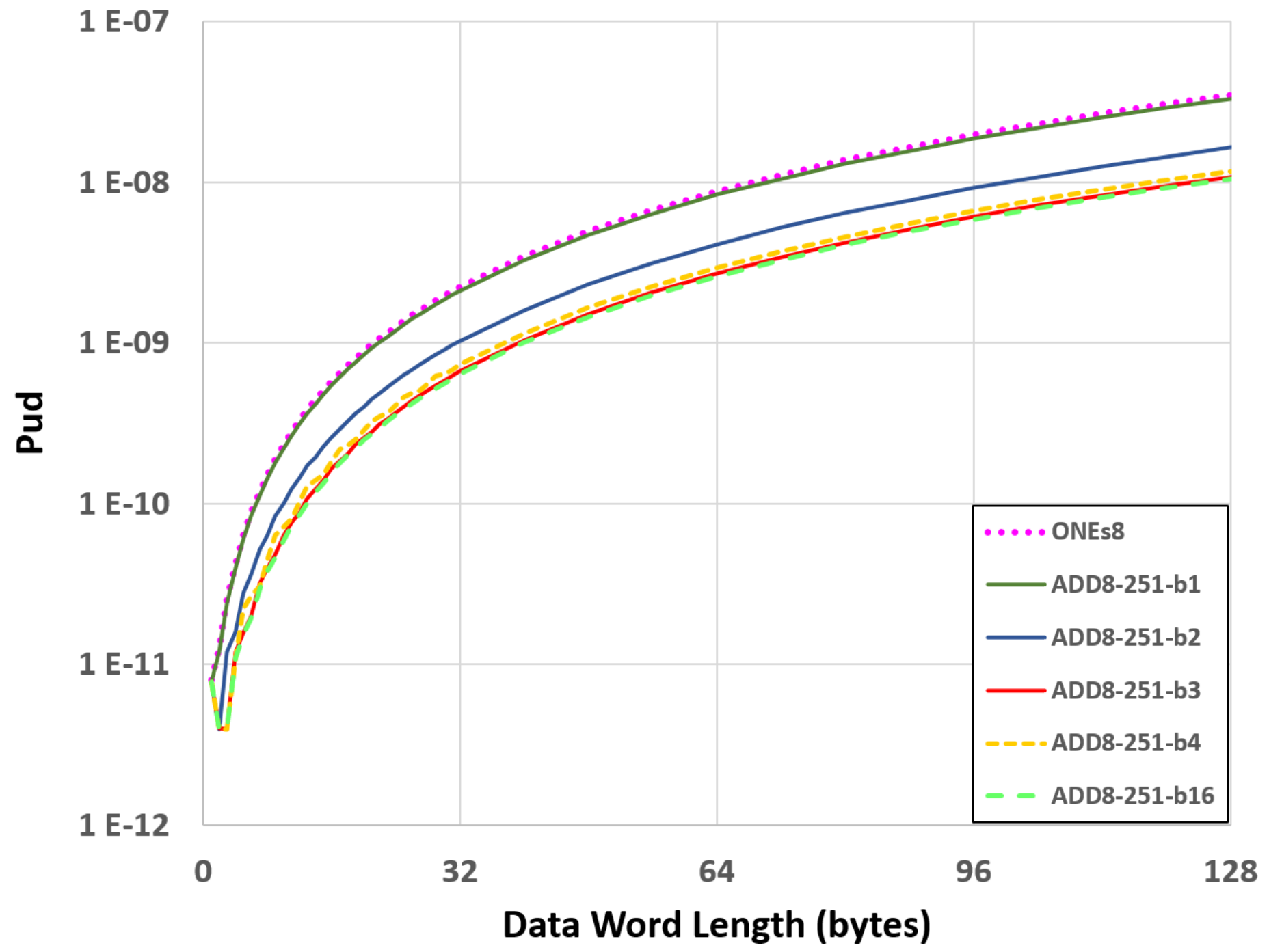

*Figure 7. 8-bit single-sum effectiveness with prime modulus 251.*

These results predict that effectiveness of modulus 255 will degrade with block size 2. They also predict that the prime modulus 251 will degrade at block size 4. On the other hand, they predict that modulus 253 will have excellent effectiveness up to a block size of 13 bytes, and modulus 239 will be good up to 14 byte blocks.

As a sanity check on these results, the following specific undetectable two-bit patterns were identified via the simulation

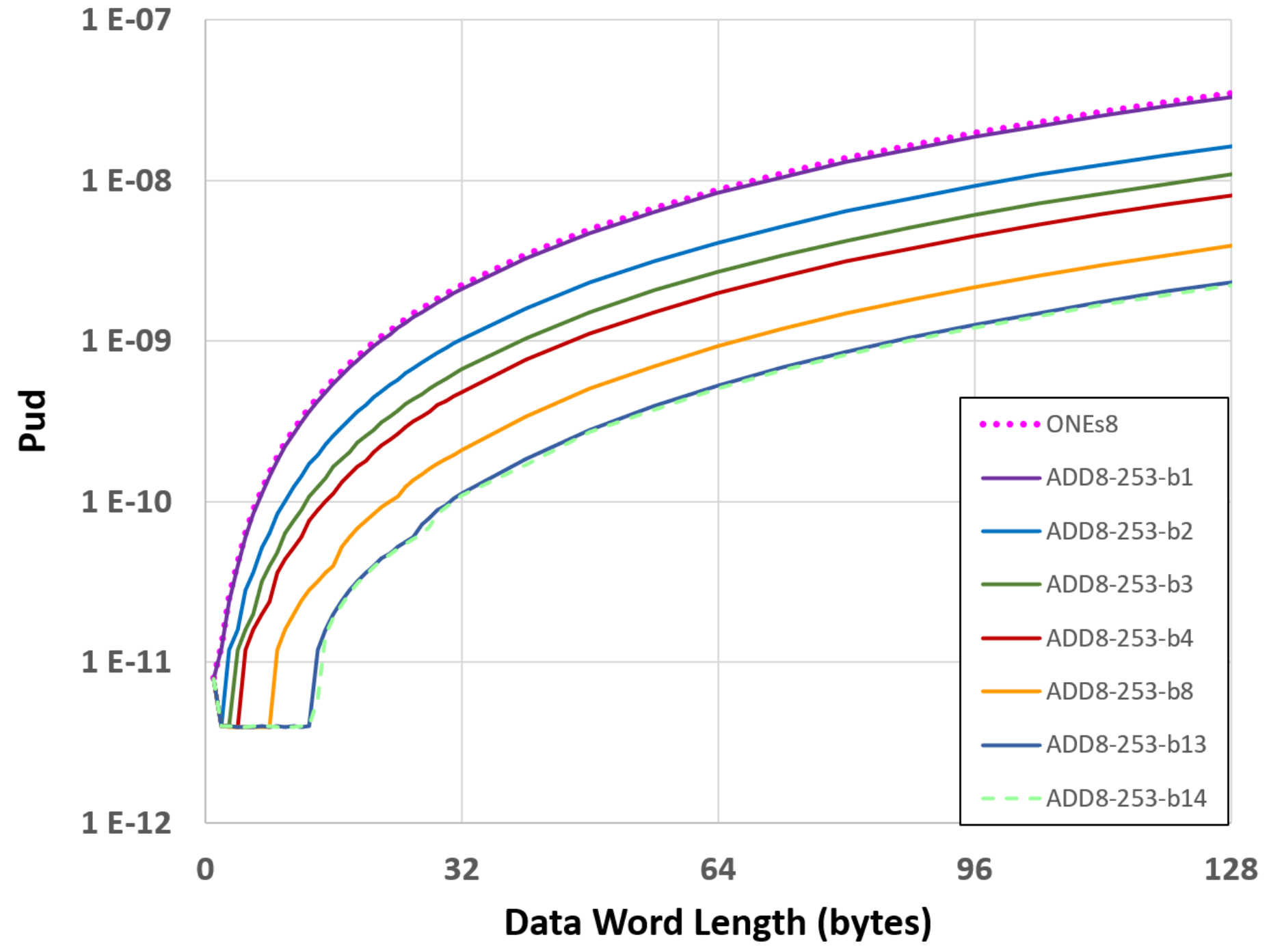

*Figure 8. 8-bit single-sum effectiveness with modulus 253.*

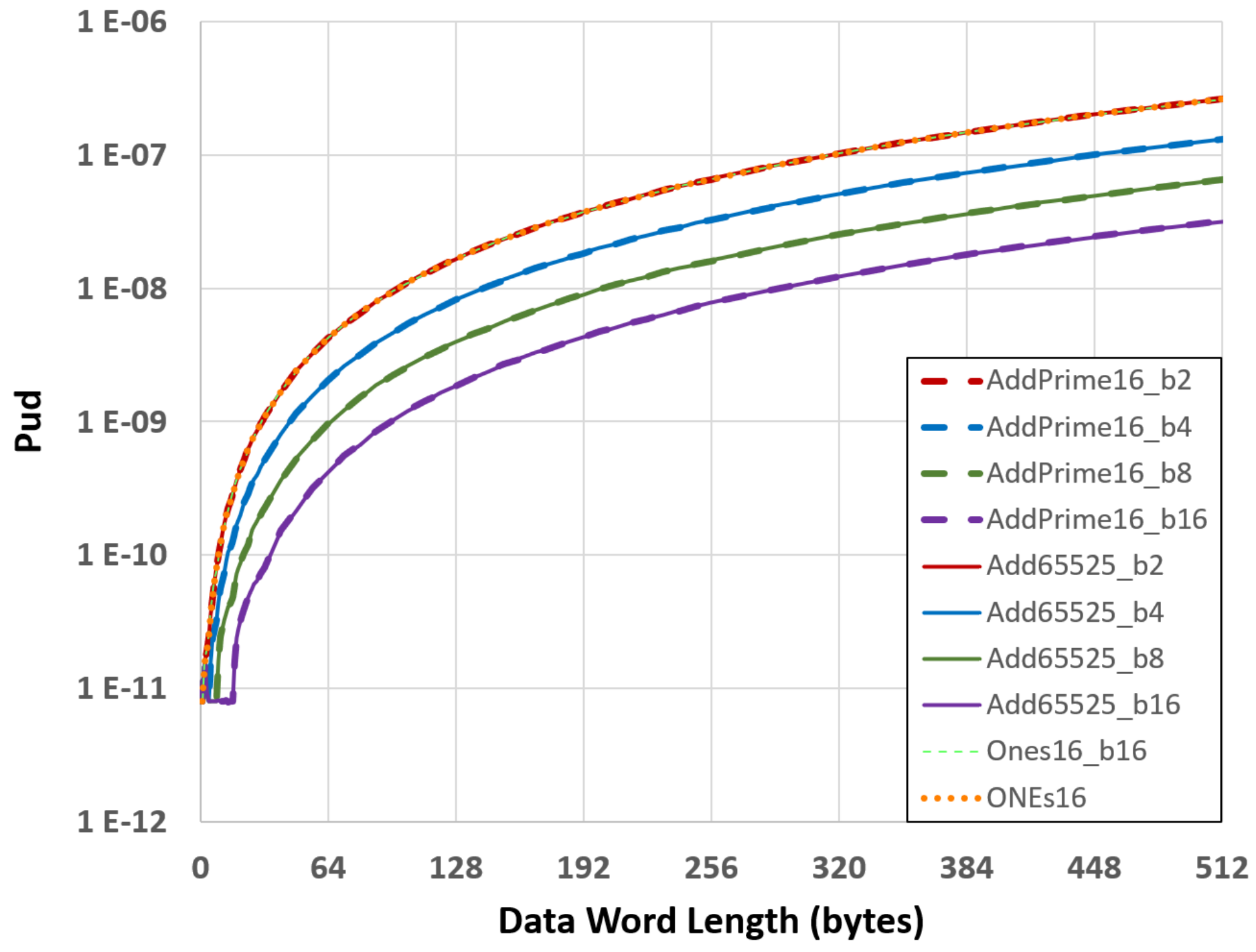

*Figure 9. 16-bit single-sum effectiveness with moduli 65521, 65525, and 65535.*

used to create figure 6. Again, the property being looked for is two block values that give the same answer (not necessarily zero) under range reduction by the modulus that differ in exactly two bit positions:

Mod 251: undetected fault at 4 byte block size

0x80000040 mod 251 = 0x00000000 mod 251 = 0

Mod 253: undetected fault at 14 byte block size

0x0000000000000000000000000002 mod 253 =
0x8000000000000000000000000000 mod 253 = 2

Offering a proof that there are no two-bit undetected faults for modulus 253 up to block sizes of 13 is beyond the scope of this paper. However, a very large number of experiments (see Section VII) failed to find a single counter-example, so we consider modulus 253 to provide resistance to two-bit faults in this manner up to a block size of 13 bytes for practical engineering purposes.

To determine the effect of this property on single-sum checksums, we examine the effectiveness of different block lengths on 8-bit checksum performance.

Figure 7 shows the significant error detection effectiveness improvement of moderately larger block sizes for modulus 251. For interpreting the legend of figure 7 and subsequent figures the notation "ADD8-251-b3" means an 8-bit single-sum checksum with modulus 251, and block size of 3 bytes.

In figure 7, a block size of 1 byte is slightly better than a one's complement checksum, as expected. However, a block size of 2 is significantly better, and a block size of 3 is better still. However, as predicted, block sizes of 4 and 16 plot on top of the block size 3 curve.

In contrast, figure 8 shows that modulus 253 provides increased effectiveness for even large block sizes. The performance of moduli 251 and 253 would be indistinguishable if the curves on figures 7 and 8 were superimposed for block sizes of 1, 2 and 3 bytes. However, modulus 253 continues to improve up to a block size of 13, stalling there with the same performance at block sizes of 14 bytes and higher. For a 128 byte data word, modulus 253 provides 14.2 times better $P_{ud}$ for 13 byte blocks than 1 byte blocks.

The wavy and horizontal portions of curves for small block sizes on figures 7 and 8 for block lengths below 32 byte data word lengths are not simulation artifacts. Those graphical features remained stable in shape while increasing the number of simulations by an order of magnitude. We believe they are caused by quantization effects for block lengths that are about the same size as, or larger than, the data word length.

Based on these results, we select modulus 253 as a promising candidate. Modulus 239 has a one-byte better range on large blocks, but has other properties for dual-sum checksums that might make it less attractive. This topic will be revisited in the context of large-block dual-sum checksums.

The error detection advantage of modulus 253 at high block lengths is maintained for all data word lengths considered (up to 4096-byte data words for this work), and there is every expectation that would continue at higher data word lengths.

*Table 3. Undetected faults for single-sum checksum with varied moduli with large block size. 128 byte data word; 2 byte check value; 16 byte block.*

| | Fraction of undetected faults | | | | |
|---|---|---|---|---|---|
| Modulus | 1-bit | 2-bit | 3-bit | 4-bit | 5-bit |
| 65535 | 0% | 3.080% | 0.2890% | 0.3124% | 0.08906% |
| 65533 | 0% | 0.343% | 0.0268% | 0.0090% | 0.00317% |
| 65531 | 0% | 0.344% | 0.0119% | 0.0067% | 0.00199% |
| 65529 | 0% | 0.343% | 0.0119% | 0.0062% | 0.00188% |
| 65527 | 0% | 0.343% | 0.0119% | 0.0058% | 0.00191% |
| 65525 | 0% | 0.343% | 0.0040% | 0.0063% | 0.00177% |
| 65523 | 0% | 0.343% | 0.0115% | 0.0085% | 0.00261% |
| **65521** | **0%** | **0.343%** | **0.0120%** | **0.0069%** | **0.00220%** |
| **65519** | **0%** | **0.343%** | **0.0120%** | **0.0059%** | **0.00196%** |
| 65517 | 0% | 0.344% | 0.0040% | 0.0061% | 0.00157% |
| 65515 | 0% | 0.343% | 0.0089% | 0.0055% | 0.00180% |
| 65513 | 0% | 0.344% | 0.0041% | 0.0049% | 0.00159% |
| 65511 | 0% | 0.344% | 0.0040% | 0.0062% | 0.00155% |
| 65509 | 0% | 0.344% | 0.0040% | 0.0055% | 0.00166% |
| 65507 | 0% | 0.343% | 0.0043% | 0.0053% | 0.00163% |
| 65505 | 0% | 0.344% | 0.0186% | 0.0126% | 0.00465% |
| 65503 | 0% | 0.343% | 0.0134% | 0.0125% | 0.00485% |
| 65501 | 0% | 0.344% | 0.0040% | 0.0050% | 0.00160% |

### *D. Large-Block 16-Bit Single Sum*

A similar situation exists for 16-bit single sums block sizes of two bytes and above. Some moduli exhibit two-bit fault vulnerabilities at increased block lengths, but a significant majority of moduli examined retain two-bit fault detection capabilities at, and likely above, 16 byte blocks.

Table 3 shows the performance of large, odd moduli for 16-bit single-sum checksums for a block size of 16 bytes and a data word size of 128 bytes. Two-bit fault detection is essentially identical except for modulus 65535, which has quite poor performance at almost nine times worse than other moduli.

At three-bit faults our preferred modulus of 65525 is three times better than the largest prime modulus of 65521, which becomes important at large data word sizes in which 3-bit faults are more likely to occur. 65525 is better than 65521 for 3-, 5-, 7-, and 9-bit faults for two data word lengths studied: 128 bytes as well as 1024 bytes.

It should be noted that the modulus matters less as the block size is reduced, with very little difference between moduli other than the one's complement modulus 65535 being a uniformly poor performer. Nonetheless, using a modulus of 253 or 65525 is always better than one's complement or prime moduli.

Figure 9 shows the comparative performance of single-sum checksums with moduli of 65521, 65525, and 65535 (one's complement). Increasing block size has no effect for modulus 65535, with block size of 2 bytes and 16 bytes both plotted on top of each other as well as the other moduli at block size 2 on the top curve. Both 65521 and 65525 have indistinguishable curves with improved effectiveness for block sizes of 4, 8, and 16 bytes.

An analogous screening process suggests the modulus 4294967283 as a good selection for 32-bit checksum addition.

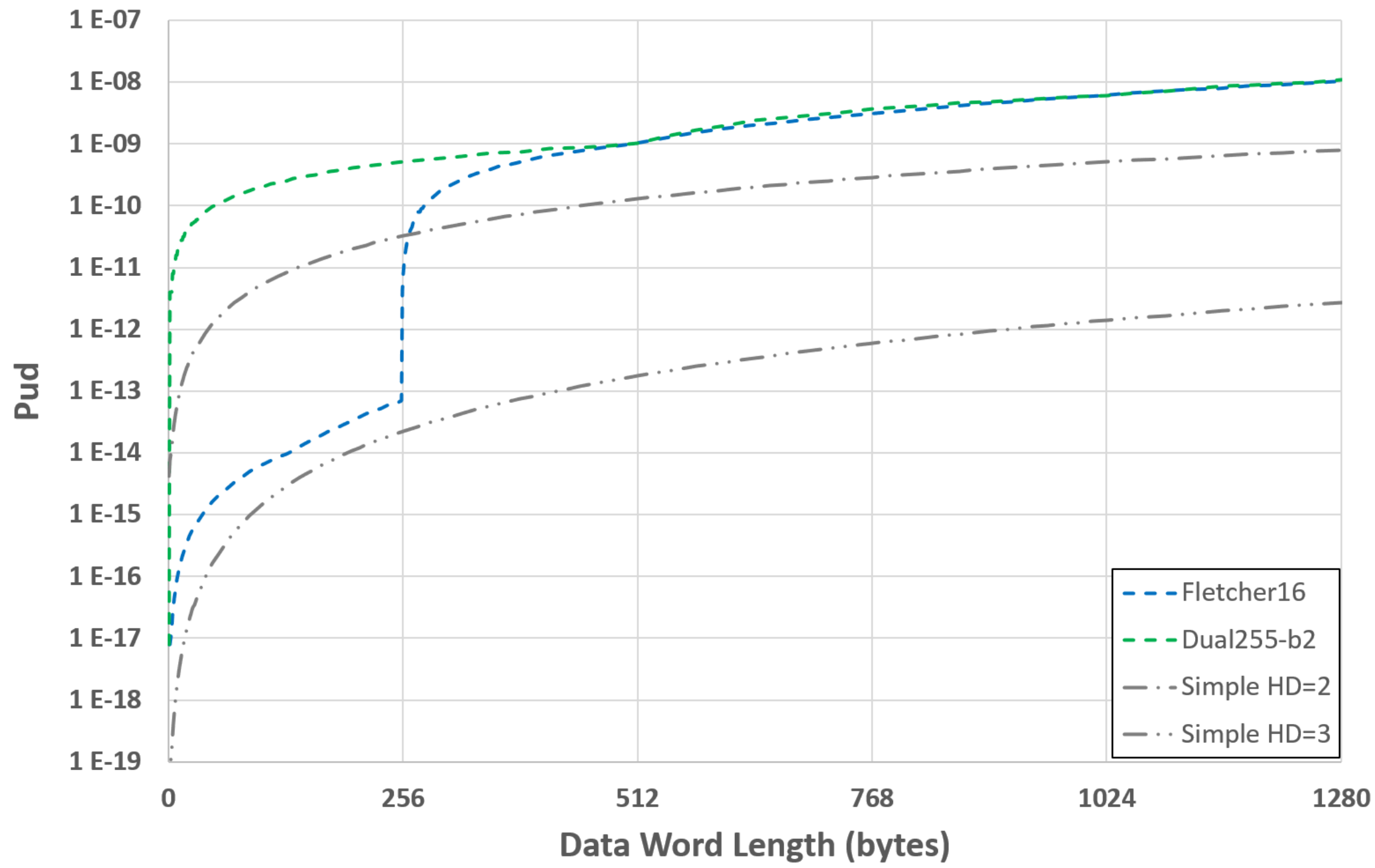

*Figure 10. Fletcher16 compared with Dual255-b2.*

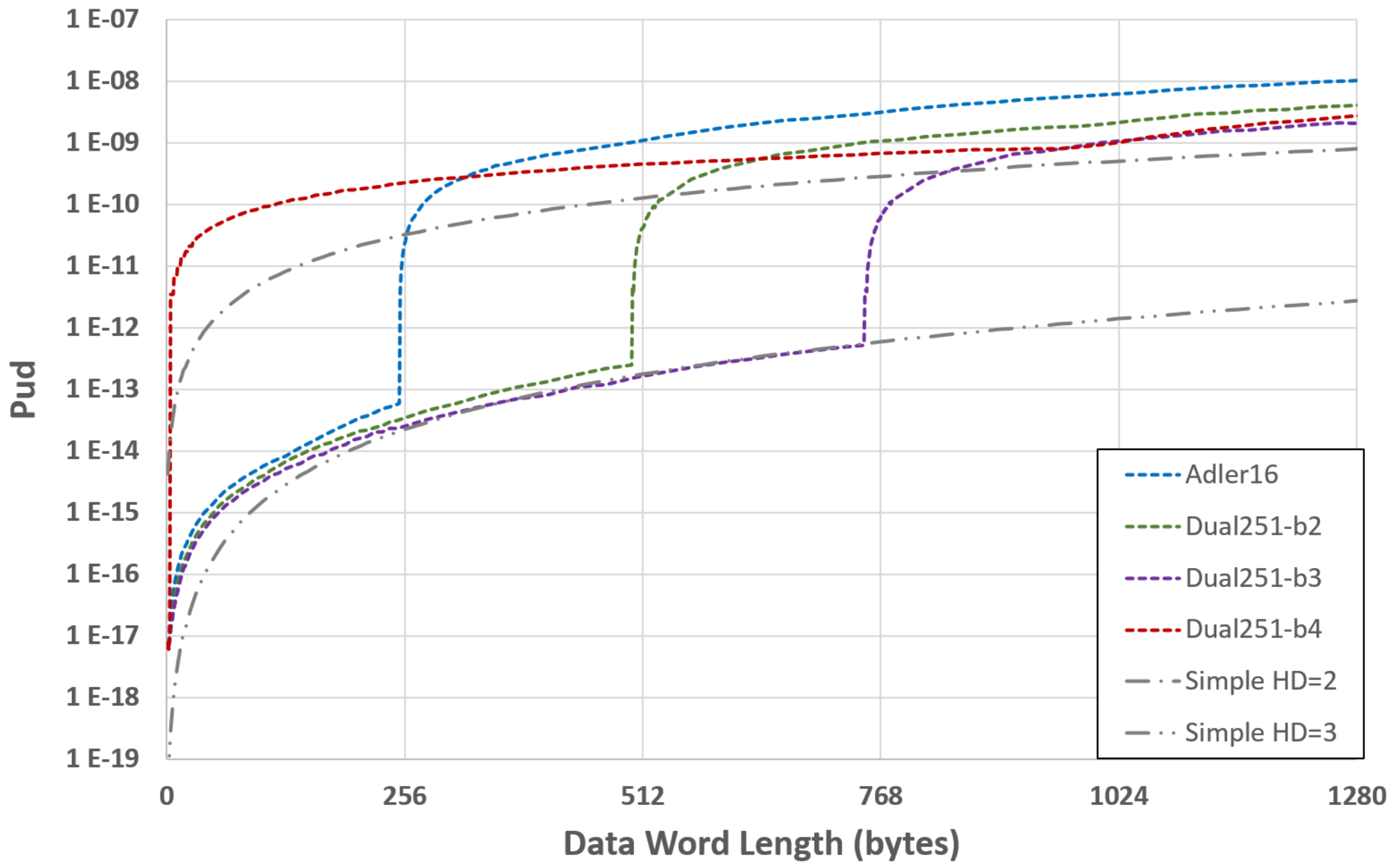

*Figure 11. Adler16 compared with longer block modulus 251 dual-sum checksums.*

## VI. Dual-Sum Large-Block Checksum Processing

Having found improvements in using a well-chosen modulus and large block sizes on single-sum checksums, we turn our attention to dual-sum checksums in the style of Fletcher and Adler checksums. (We refer to the extended block versions as one's complement and largest prime dual-sum checksums since large-block performance was not a design objective of those earlier works. Any comparative lack of effectiveness at large block sizes was beyond the scope of the designers who created those checksum algorithms, and should not be considered criticism of their work.)

### A. *One's Complement Dual-Sum Modulus*

As was the case for single-sum checksums, using a one's complement modulus provides no benefit with increased block size for a dual-sum checksum algorithm. Indeed, the poor performance of modulus 255 becomes apparent at a block length of 2, just as was seen with modulus 255 for single-sum checksums. Figure 10 shows a Fletcher16 checksum (dual 8-bit sums with modulus 255) with block length of one and block length of two ("Dual255-b2"). At a block length of 1, the HD=3 capability is 254 data word bytes. At block length of 2, it reduces to 1 byte.

Figure 10 also reveals a more subtle feature of dual-sum checksum performance, which is that there are multiple effectiveness inflection points. The first inflection point is the obvious HD=3 capability at data word length 254 bytes for a Fletcher16 checksum. The second is at 508 bytes, which is double the rollover length of the SumB variable. The 2-byte block effectiveness converges with the 1-byte block curve in the vicinity of this second inflection point.

### B. *Largest Prime Dual-Sum Modulus*

A largest-prime modulus creates a striking effect with increased block sizes by doubling or tripling the length of the HD=3 capability.

Figure 11 shows that a largest-prime modulus of 251 has HD=3 capability of 250 bytes as expected. Moreover, increasing the block length increases that capability to 500 bytes (2 byte blocks), and 750 bytes (3 byte blocks). A block size of 4 bytes "breaks" the dual-sum approach with this modulus, with an HD=3 capability of 3 data word bytes.

To be clear about the implications of this graph: computing an Adler16 checksum using two bytes of the data word at a time (block size of 2 bytes) increases the HD=3 capability from 250 bytes to 500 bytes, with no other change to the checksum algorithm beyond increased block size. A block size of 3 bytes extends the HD=3 capability to 750 bytes. Beyond that, larger block sizes degrade effectiveness severely to be HD=2 at all but the very shortest data word lengths. As with modulus 255, the high-block-length curves merge with the highest good HD block size at about one rollover length beyond the HD=3 capability of the largest good block length.

### C. *A Better Dual-Sum Modulus*

To be sure, tripling the HD=3 capability of an Adler checksum by increasing the block size (checksum algorithm Dual251-b3) is impressive. But this immediately raises two questions: why did Dual251-b4 fail so dramatically? And, is there a way to do even better?

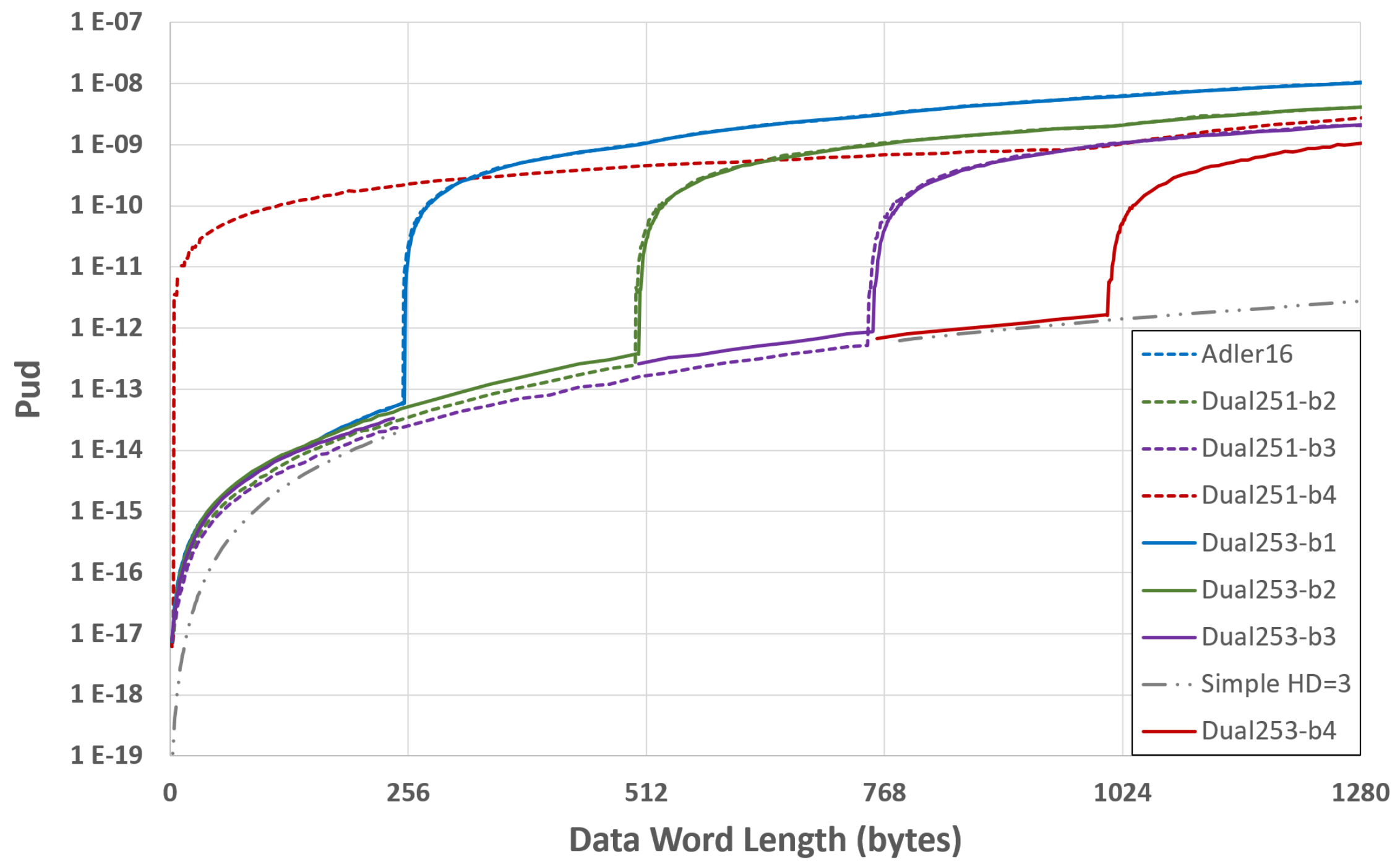

*Figure 12. Modulus 251 vs. Modulus 253 for dual-sum checksums. Portions of Dual253-b3 and Dual253-b4 curves omitted for clarity.*

The reason that Dual251-b4 failed so dramatically can be seen in Figure 6 in the discussion on single-sum checksums. A modulus of 251 is vulnerable to two-bit undetected errors at a block size of 4, and we saw Dual251-b4 fail.

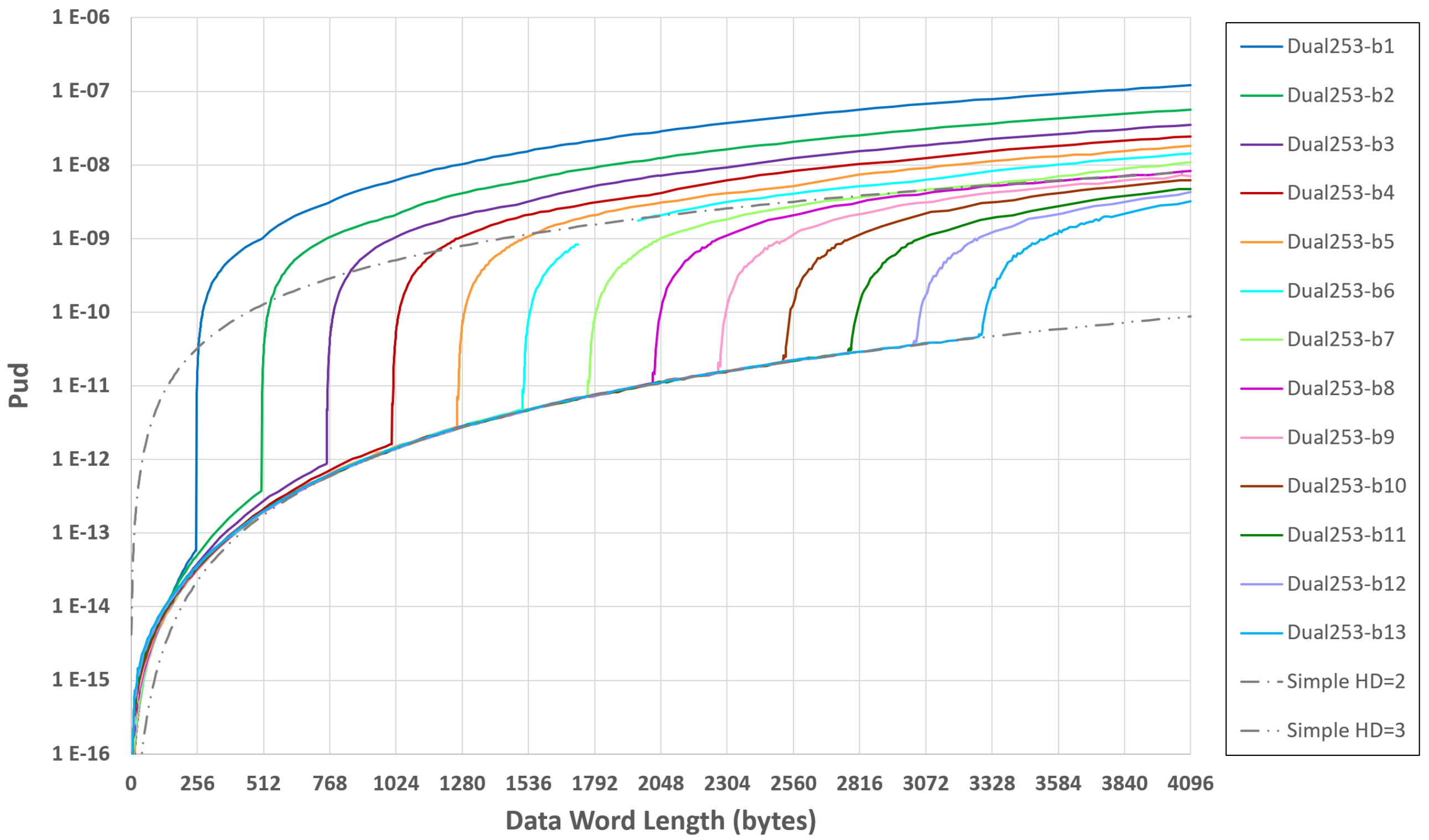

*Figure 13. Large block 16-bit dual-sum checksum performance. Modulus 253; block sizes range from 1 to 13 bytes.*

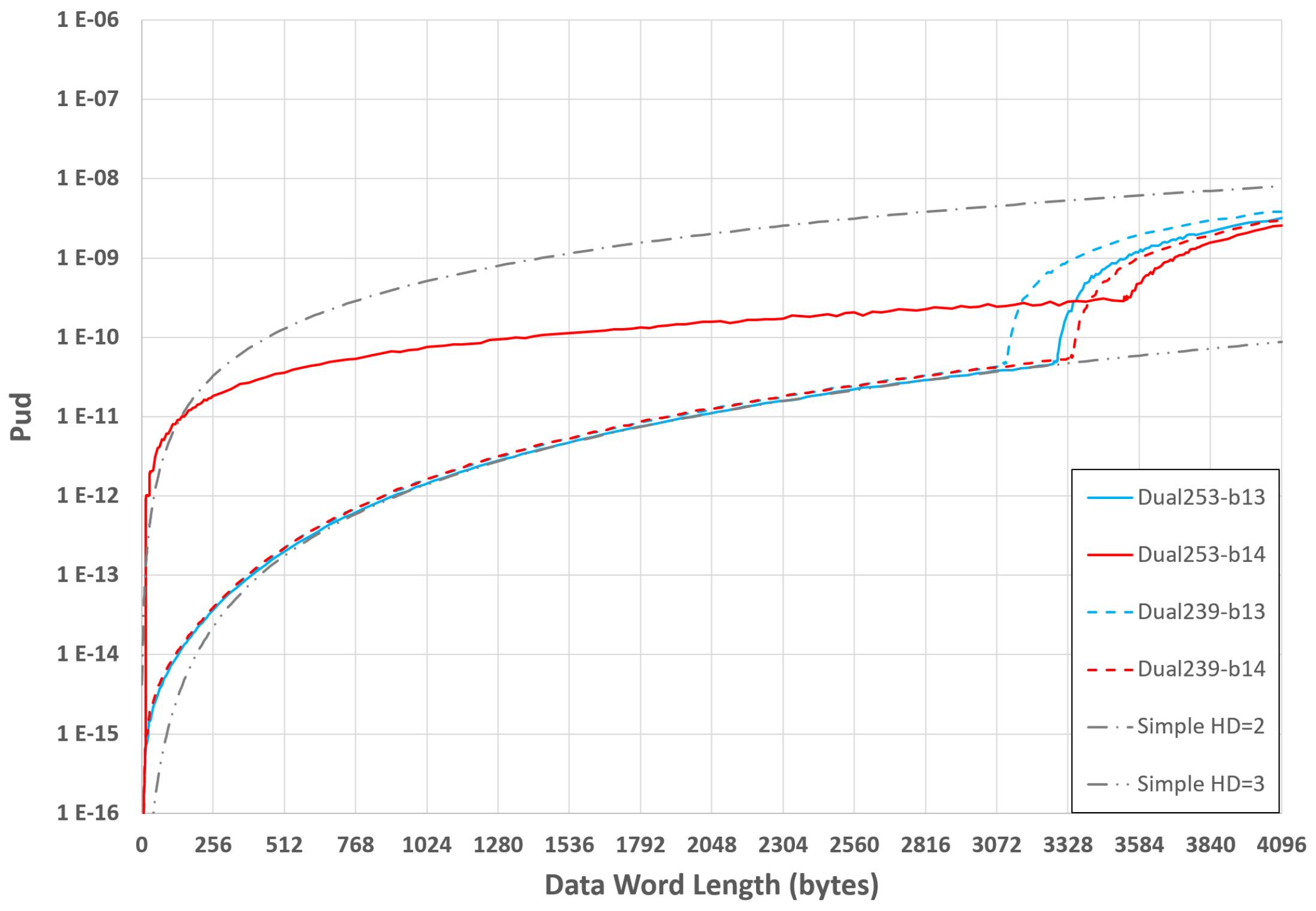

*Figure 14. Modulus 253 vs. Modulus 239 long-block dual-sum checksum performance comparison.*

Still referring to Figure 6, modulus 253 looks much better. And it turns out it is also better for dual-sum checksum use, providing excellent results through Dual253-b13.

Figure 12 shows that Dual253 checksums perform almost identically to Dual251 checksums for a block size 1 byte. At a block size of 2 bytes, Dual251-b2 has a slightly better effectiveness, but a slightly shorter HD=3 capability due to the modulus being different. The difference increases at a block length of 3. (Portions of Dual253-b3 and Dual253-b4 curves have been omitted from the graph to be able to see Dual251-b2 and Dual251-b3 curves clearly.)

However at a block size of 4, Dual251-b4 switches to HD=2 behavior for any but the smallest block sizes, while Dual253-b4 moves the HD=3 capability by yet another multiple of its rollover length.

It is also interesting to note that both dual-sum checksums get closer to the Simple HD=3 trend line as block size increases. This appears to be due to improved bit mixing of the modulus operation on larger blocks, with the modulus operation increasingly approximating a randomized check value at HD=3.

This improved block size capability for modulus 253 is not a fluke, but related to the vulnerability to two-bit faults in the modulo operation previously discussed. Exactly as predicted, figure 13 shows that Dual253 checksums provide improved HD=3 performance up to block sizes of 13 bytes, with the HD=3 capability increases by 252 data word bytes for each one-byte increase in block size. Modulus 253 reverts to HD=2 behavior at 14 bytes (figure 14).

$P_{ud}$ degrades quickly above the HD=3 capability as with other dual-sum checksums. However, in an echo of performance with single-sum checksums, longer block lengths maintain better HD=2 performance past their HD=3 capability due to improved large-block bit mixing.

Even a single two-bit undetected fault makes a bigger contribution to $P_{ud}$ at a modest BER due to the much higher probability of two-bit faults compared to three-bit faults. This accounts for the steep slope of the $P_{ud}$ curve at the HD=3 capability. After that, the initially small fraction of undetected two-bit faults increases to the point that contributions from undetected three-bit faults no longer dominate the $P_{ud}$ result.

*Table 4. Observed HD=3 capability for different dual-sum checksum moduli.*

| Modulus | Block Size | HD=3 capability |
|---|---|---|
| 255 | 1 | 254 |
| 255 | 2 | 1 |
| 251 | 1 | 250 |
| 251 | 2 | 500 |
| 251 | 3 | 750 |
| 251 | 4+ | 3 |
| 253 | 1 | 252 |
| 253 | 2 | 504 |
| 253 | 3 | 756 |
| 253 | … | block size * 252 |
| 253 | 12 | 3024 = 12*252 |
| 253 | 13 | 3276 = 13*252 |
| 253 | 14+ | 13 |
| 239 | 13 | 3094 = 13*238 |
| 239 | 14 | 3332 = 13*238 |
| 239 | 15+ | 14 |

As expected, Dual253-b14 has poor effectiveness, with an HD=3 capability of 13 bytes. (This behavior of still being good up to 13-byte data words helps confirms the analysis discussed in the next section.)

### *D. What About Modulus 239?*

Figure 6 showed that modulus 239 had even better long-block capability than modulus 253, by one byte of block length. However, the HD=3 capability of any given block length happens at shorter lengths (multiples of 238 bytes, rather than multiples of 252 bytes).

Figure 14 shows a comparison of 13 and 14 byte data word lengths for moduli 239 and 253. The expected behavior of a HD=3 capability up to 14 byte block length for modulus 239 is confirmed. But due to the shorter rollover length, the difference between 252*13 and 239*14 maximum HD=3 capabilities is so small that there is only a very narrow window of data word size that favors modulus 239. (Omitted from the graph for clarity is confirmation that indeed modulus 239 degrades to HD=2 behavior for a block size of 15 bytes as expected.)

Based on these results, we recommend that modulus 253 with the largest block size reasonably supported by the implementation platform be used as a general-purpose dual-sum checksum.

## VII. Fault Detection Performance

### *A. Maximum effective block size*

The HD=3 capabilities of large-block dual-sum checksums provide a pattern that helps understand how they work. Each modulus provides HD=3 at increasing data word lengths with increased block size – up to a point. Beyond that point effectiveness degrades to HD=2 at very short data word lengths.

Table 4 summarizes data for the four moduli studied for dual-sum checksums. The observed data reveals the pattern:

$$HD3Capability = BlockSize * (Modulus-1) \qquad (5)$$

Intuitively the size of the sum might seem like it would have an effect for moduli larger than one byte. However, this is negated by the need to divide the block into larger pieces to feed the sum. So a 16-byte block provides a 16:1 digest size compression for 1-byte sums, but only an 8:1 digest size compression when converting 16-byte blocks into 2-byte sum inputs. Thus, the size of the sum cancels out and is not included in the formula.

This empirical relationship of HD=3 capability to block size and modulus permits us to assemble an explanation of the large block error detection mechanism for dual-sum checksums.

### *B. Large Block Error Detection Mechanism*

Before relying on the effectiveness of large block checksums, it is reasonable to want to understand the mechanism that provides their advantages. It involves a combination of the dual-fault resistance properties of the chosen modulus, plus the dual-sum checksum rollover length, both of which have been previously described.

Similar to our earlier discussion of undetected fault mechanisms for single-sum checksums, there are three ways for a two-bit fault to be undetected in a dual-sum checksum:

1. A one bit fault in the check value that compensates for a one-bit fault in a data word. However, for split sum checksums these can only occur in data words larger than the rollover length of the modulus.
2. Two separate bit faults that are separated in the data word by at least the rollover length.
3. Two separate bit faults within a single block that result in an unchanged modulus operation result, presenting an unchanged input to the summing operation.

The third mechanism becomes problematic for large-block checksums, as was seen for single-sum checksums.

These fault mechanisms account for long-block dual-sum checksum fault behavior if the range reduction function of the modulus is considered as computing a sequence of digest values that are then fed into the dual-sum checksum.

Figure 15 shows a conceptual diagram of how a long-block checksum works. The concept is general to both single-sum and dual-sum checksums, but a specific example has been chosen for illustrative purposes of a modulus 253, block size 8 byte, and a 2-byte check value for a dual-sum checksum, which will achieve an HD=3 capability for a 2016 byte data word.

In figure 15, the data word of 2016 bytes is broken down into 252 blocks for processing, with each block being 8 bytes in size (a Dual253-b8 checksum algorithm in this example). Each block of 8 bytes undergoes a range reduction via a mod 253 operation, resulting in a one byte remainder value that is, for practical purposes a digest (or hash value) of the corresponding 8-byte block. This results in a sequence of 252 digest bytes. Those digest bytes are then processed by a dual-sum checksum that also uses mod 253 to produce a check value.

Because the modulus 253 is resistant to undetectable two-bit faults as a digest function at and below 13-byte blocks, we know that any 1- or 2-bit fault in a single block will produce at least a one-bit change in its digest compared to the fault-free digest value. In other words, if the block has a 1- or 2-bit fault, the digest will also have some fault.

We also exploit a property of dual-sum checksums not previously discussed. Dual-sum checksums do better than providing HD=3 up to one less than the modulus size in bytes. They provided detection of any two corrupted inputs to the dual-sum process even if more than two bits have been corrupted within those two inputs – so long as those two inputs are separated by less than the HD=3 capability of the checksum. (This property is a consequence of the dual-sum approach; resistance to two-bit faults with one bit inverted in each of two data words is simply a special case of this property.) Thus, regardless of the number of bits changed in the digest of two blocks, as long as only two digest bytes are corrupted, a dual-sum checksum addition scheme will detect them.

The same analysis applies to dual-sum checksums of other sizes. For example, a 32-bit dual-sum checksum using a modulus of 65525 is expected to have the same properties for the same reasons up to at least a block size of 16 bytes (8 times the sum-size of 2 bytes).

*C. Approximately HD=3*

Because this is an empirical, simulation-based approach, we cannot quite claim that large block checksums absolutely give HD=3 for longer data word lengths than checksums with a block size the same as the modulus size. Rather, they appear to give HD=3 per the mechanisms described in the preceding subsection.

The missing link for a true guarantee is proving that there is no undetectable two-bit fault that will slip past the modulus operation being used as a digest function. In other words, equality (4) should be proven to have no solution for the maximum block size and modulus being used.

In practice, the $P_{ud}$ graphs make it clear that large-block checksums have the properties found for each modulus identified *as far as we know*. Moreover, the number of simulations run shows that even if some undetectable two-bit faults were to exist in principle in a single block, they are so rare that they make no practical difference under a random independent bit inversion fault model.

To increase informal confidence, we ran the fault injection simulation used to create figure 6 for additional experiments. Trillions of random trials failed to find an undetected two-bit corruption of random data words missed by moduli 253 at a 13-byte block length, or 65525 at a block length of 16 bytes. So we consider these moduli to be close enough for practical purposes to supporting a digest function that provides HD=3-class performance at the stated block lengths.

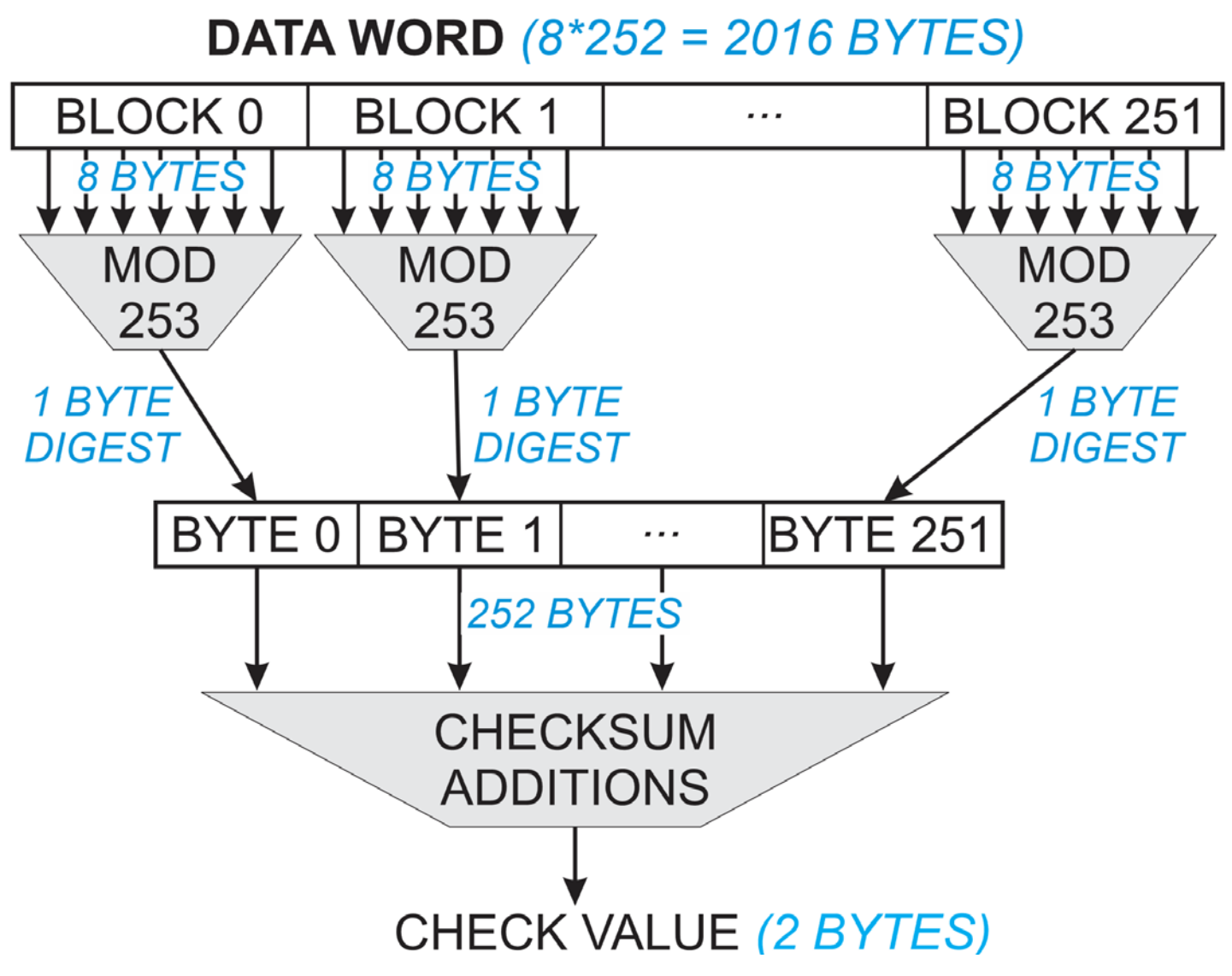

*Figure 15. Conceptual diagram of operation of a large-block dual-sum checksum with block size 8 bytes and modulus 253.*

*D. Implementation Considerations*

The large-block checksum algorithm's effectiveness does not depend on byte ordering or endian-ness of the data word for error detection effectiveness. Similarly, byte ordering in the check value does not matter for effectiveness. All that matters is that the software constructing the initial code word has compatible ordering with the software checking the code word integrity.

There is a potential accuracy issue if the block size is the same size as an architected register size. Adding a new block to a running sum can result in a carry-out of the sum, which might be lost on computers that do not have an architected carry bit visible to the programmer.

One solution to avoiding a lost carry-out is to use a higher precision intermediate sum if that is conveniently available in the programming language being used. For example for an 8-byte block, a modular addition might be carried out as a 128-bit addition, followed by a modulo operation that takes a 128-bit input and produces a smaller output appropriate for the sum size. Using C type-casting notation with a 32-bit running sum and a 128-bit variable holding a full block, that looks like

```
sum32 = (uint32_t) ( ( ((uint128_t) sum32) + block128) %
                                          modulus32 );
```

This creates a 128-bit unsigned sum and produces a 32-bit result of a modulo reduction. We note that "uint128_t" is non-standard notation, but is intended to represent a casting to a 128 unsigned integer in the manner of "cstdint" notation which defines uint32_t as the type of an unsigned 32-bit integer [C++].

Another approach is to perform range reduction before the sum, which might be faster on some computers. That option looks like this:

```
block32 = (uint32_t) (block128 % modulus32);
sum32  = (sum32 + block32) % modulus32;
```

This option requires two modulo operations, but the second modulo operation can be faster since it only operates on 32 bits instead of a larger block size. There is still the possibility of a carry-out from the addition, so this approach only works if the modulus is at least one bit smaller than the sum size (e.g., a modulus less than 2**31 for 32-bit unsigned integers). A 32-bit register approach with this technique will work well for an 8-bit or 16-bit sized modulus for example (moduli 253 or 65525).

See Appendix A for some example variations to make this discussion more concrete. The examples in Appendix sections D & E additionally show variations that delay applying the modulus operation until the end of the checksum for speed, although care must be taken to ensure the data word is short enough to avoid sum overflows. These techniques can dramatically reduce the cost of applying the modulus.

It is important to ensure that all variables are declared as "unsigned" to avoid inadvertent sign extension while assembling blocks from the data word and performing the modulo operation properly with maximum-size sums. Using appropriate type casting is both tricky and critical to achieving correct results.

The net speed of a large block checksum depends on the implementation, but can potentially be faster than for small-block checksums when run on architectures that have large register sizes. While a modulus must be computed for each summed block, those blocks are processed in larger chunks, with only one division operation per block regardless of the block size. Those divisions can run comparatively fast compared to memory access overhead if the block fits within a single hardware register. Even with less capable machines, the cost of division is linear with the size of the dividend, so fewer large divisions can take no more net time than a larger number of small divisions operating on the same data word size.

Other speed-up tricks can also be used, such as delaying the modulo operations until after several additions have been performed, especially on SumB, so long as integer registers are not overflowed.

## VIII. Conclusions

This paper makes the following contributions:

1. An exploration of modular addition operations for checksums reveals that the best modulus might not be any of the two's complement, one's complement, or largest prime moduli – but rather an empirically validated odd modulus that just happens to give good error detection performance. The key to this approach is finding a modulus that does not let two-bit faults become undetectable in a remainder-from-division digest computation. The moduli 253 and 65525 are identified as having good single-sum checksum performance for 8- and 16-bit checksums respectively.

2. Increasing the block size processed by a modular checksum operation is identified as significantly improving error detection capabilities in both single-sum and dual-sum checksum algorithms, so long as a suitable modulus is used. Again, moduli 253 and 65525 are recommended.

3. A carefully selected modulus can provide significantly higher HD=3 capability for dual-sum checksum algorithms. The modulus 253 is identified as providing HD=3 performance for a large-block version of the Fletcher/Adler dual 8-bit checksum algorithm using a modulus of 253 up to a data word size of 504 for 2-byte blocks, 1208 for 4-byte blocks, and a maximum 3276 bytes for 13-byte blocks. A dual checksum using modulus 65525 looks good, but was only partially validated due to limited computational resources, with apparent HD=3 capability up to at least 16*65524 data word bytes, which is just under a one megabyte data word size. (HD=3 might be maintained at longer data word lengths, but that is beyond the ability of the experimental framework available for this work to validate.)

4. The mechanism for why large-block checksum algorithms are effective is explained. In short, the modular addition computes a digest of a large block before combining it with a running sum. This preserves the HD=3 capability of a dual-sum checksum for a multiple of the normal size determined by the ratio of the block size to the sum size. (E.g., a 12-byte block with a 32-bit dual-sum checksum breaks each 12-byte block into six two-byte values for the rolling sum, giving a factor of 6 increase in performance of the HD=3 fault detection capability.)

It is important to remember that Cyclic Redundancy Checks (CRCs) can provide far superior fault detection mechanisms to even these improved checksum approaches. Nonetheless, using better moduli and longer block sizes can dramatically improve error detection effectiveness with the same check value size and comparable computational cost compared to previously known checksum approaches.

## IX. Acknowledgements

The author wishes to thank Theresa Maxino for her collaboration on an earlier generation of this research.

## APPENDIX A: EXAMPLE INNER LOOP CODE

The below C code fragments are intended to illustrate the key idea behind the use of large-block checksums. They are written in a way to make the key ideas obvious. They are not intended as an illustration of portability or otherwise-desirable code structure.

Code fragments assume without checking that the data word is evenly divisible by the block size, or has been zero-padded to be so to keep the example simple. Variable typing is per <cstdint>, <stdint.h>, or a similar definition approach.

A data word organized as a sequence of bytes is assumed to be in dataWord8 as a uint8_t data array, whereas dataWord32 is a uint32_t data array that holds 32 bits of the data word in each array element. The endian-ness of assembling bytes into a block in the byte-organized example will not affect checksum effectiveness. The variable dwSize is assumed to be the number of relevant elements in the data word array.

All code examples will provide HD=3 capability up to a data word size of 252 data elements of 4-byte blocks = 252*4 ➔ 1008 bytes. Note that examples D and E defer the modulus operation for speed. Example E makes an assumption that the data word size is small enough to avoid sumA overflow.

*A. Dual-Sum, Block Size 4, byte-organized data word*

```
uint32_t Dual16b4_A(uint8_t dataWord8[],
                                        uint32_t dwSize)
{ uint32_t sumA = 0;
  uint32_t sumB = 0;
  for( uint32_t index = 0; index < dwSize; index += 4 )
  { uint32_t block =   (dataWord8[index])
                 | (dataWord8[index+1] << 8)
                 | (dataWord8[index+2] << 16)
                 | (dataWord8[index+3] << 24);
    uint32_t digest = block % 253;
    sumA = (sumA + digest) % 253;
    sumB = (sumB + sumA) % 253;
  }
  return(sumA | (sumB << 8));
}
```

*B. Dual-Sum, Block Size 4, 32-bit-organized data word*

```
uint32_t Dual16b4_B(uint32_t dataWord32[],
                         uint32_t dwSize)
{ uint32_t sumA = 0;
  uint32_t sumB = 0;
  for( uint32_t index = 0; index < dwSize; index += 1)
  { uint32_t digest =  dataWord32[index] % 253;
    sumA = (sumA + digest) % 253;
    sumB = (sumB + sumA) % 253;
  }
  return(sumA | (sumB << 8));
}
```

*C. Alternate Dual-Sum, Block Size 4, 8-bit-organized data word*

```
uint32_t Dual16b4_C(uint8_t dataWord8[],
                                        uint32_t dwSize)
{ uint64_t sumA = 0;
  uint64_t sumB = 0;
  for( uint32_t index = 0; index < dwSize; index += 4 )
  { uint64_t block =   ((uint64_t)dataWord8[index])
                 | ((uint64_t)dataWord8[index+1] << 8)
                 | ((uint64_t)dataWord8[index+2] << 16)
                 | ((uint64_t)dataWord8[index+3] << 24);
    sumA = (uint32_t)((((uint64_t) sumA) + block) % 253);
    sumB = (sumB + sumA) % 253;
  }
  return((uint32_t)sumA | ((uint32_t)sumB << 8));
}
```

*D. Alternate Dual-Sum, Block Size 4, 32-bit-organized data word, delayed modulo sumB for speed*

```
uint32_t Dual 16b4_D(uint32_t dataWord32[],
                           uint32_t dwSize)
{ // Beware overflow of sumB for dwSize > 2**24
  uint32_t sumA = 0;
  uint32_t sumB = 0;
  for(uint32_t index = 0; index < dwSize; index += 1)
  { sumA = (uint32_t) (((uint64_t)sumA
            + (uint64_t) dataWord32[index]) % 253);
    sumB = (sumB + sumA);
  }
  sumB = sumB % 253;
  return(sumA | (sumB << 8));
}
```

*E. Alternate Dual-Sum, Block Size 4, 32-bit-organized data word, delayed modulo sumA and sumB for speed*

```
uint32_t Dual 16b4_E(uint32_t dataWord32[],
                           uint32_t dwSize)
{ // Beware overflow of sums for large dwSize
  uint64_t sumA = 0;
  uint64_t sumB = 0;
  for(uint32_t index = 0; index < dwSize; index += 1)
  { sumA = sumA + (uint64_t) dataWord32[index];
    sumB = sumB + sumA;
  }
  sumA = sumA % 253;
  sumB = sumB % 253;
  return((uint32_t)sumA | ((uint32_t)sumB << 8));
}
```